\documentclass[letterpaper, 10 pt, conference]{ieeeconf} %\documentclass[a4paper, 10pt, conference]{ieeeconf}      % Use this line for a4 paper

\IEEEoverridecommandlockouts                              % This command is only needed if 
\usepackage{graphics} % for pdf, bitmapped graphics files
\usepackage{epsfig} % for postscript graphics files
\usepackage{subfigure}
\usepackage{epstopdf}
\usepackage{times} % assumes new font selection scheme installed
\usepackage{amsmath} % assumes amsmath package installed
\usepackage{amssymb}  % assumes amsmath package installed
\usepackage{hyperref} 
\usepackage{color}
\usepackage[T1]{fontenc}
\usepackage{algorithm}
\usepackage{algpseudocode}
\usepackage{multirow}
\usepackage{ctable} % for \specialrule command
\usepackage{caption}
\usepackage{cite}
\usepackage{amsfonts}
\usepackage{comment}
\usepackage{footnote}

\title{\bf Cloud-Assisted Nonlinear Model Predictive Control \\ for Finite-Duration Tasks}

\author{Nan Li$^{*}$, Kaixiang Zhang$^{*}$, Zhaojian Li$^{**}$, Vaibhav Srivastava, Xiang Yin
\thanks{$*$ The authors contribute equally to this paper.}
\thanks{$**$ Zhaojian Li is the corresponding author.}
\thanks{Nan Li is with the Department of Aerospace Engineering, University of Michigan, Ann Arbor, MI 48109, USA (e-mail: nanli@umich.edu).}
\thanks{Kaixiang Zhang and Zhaojian Li are with the Department of Mechanical Engineering, Michigan State University, East Lansing, MI 48824, USA (e-mail: zhangk64@msu.edu, lizhaoj1@egr.msu.edu).}
\thanks{Vaibhav Srivastava is with the Department of Electrical and Computer Engineering, Michigan State University, East Lansing, MI 48824, USA (e-mail: vaibhav@egr.msu.edu).}
\thanks{Xiang Yin is with the Department of Automation and Key Laboratory of System Control and Information Processing, Shanghai Jiao Tong University, Shanghai 200240, China (e-mail: yinxiang@sjtu.edu.cn).}
}

\begin{document}

\maketitle
\thispagestyle{empty}
\pagestyle{empty}

\begin{abstract}
Cloud computing creates new possibilities for control applications by offering powerful computation and storage capabilities.  In this paper, we propose a novel cloud-assisted model predictive control (MPC) framework in which we systematically fuse a cloud MPC that uses a high-fidelity nonlinear model but is subject to communication delays with a local MPC that exploits simplified dynamics (due to limited computation) but has timely feedback. Unlike traditional cloud-based control that treats the cloud as powerful, remote, and sole controller in a networked-system control setting, the proposed framework aims at seamlessly integrating  the two controllers for enhanced performance. In particular, we formalize the fusion problem for finite-duration tasks by explicitly considering model mismatches and errors due to request-response communication delays. We analyze stability-like properties of the proposed cloud-assisted MPC framework and establish approaches to robustly handling constraints within this framework in spite of plant-model mismatch and disturbances. A fusion scheme is then developed to enhance control performance while satisfying stability-like conditions, the efficacy of which is demonstrated with multiple simulation examples, including an automotive control example to show its industrial application potentials.
\end{abstract}

%\IEEEpeerreviewmaketitle

\section{Introduction}\label{sec:intro}

Cloud computing is a computing paradigm that has evolved significantly over the past few years. In general, cloud computing is the availability of computer system resources to provide on-demand computing power and data storage services to users \cite{grossman2009case}. The ``infinite'' computing capacity of the cloud has opened up new possibilities for industrial automation \cite{givehchi2013cloud,hegazy2014industrial} and control applications \cite{givehchi2014control,li2014cloud,li2015road,li2016road}, especially for optimization-based and learning-based control strategies that are computation and/or data intensive \cite{camacho2007mpc,alpaydin2020introduction}. Meanwhile, moving onboard computations to the cloud has given rise to new problems related to communication delays \cite{mubeen2017delay,pelle2019towards}, connection and/or packet losses \cite{skarin2019cloud,skarin2020cloud}, as well as privacy and cybersecurity issues \cite{darup2020encrypted}.

The majority of existing work on cloud-based control systems has focused on high-level abstraction and design of computation and control architectures \cite{chen2010robot,delsing2012migration,sequeira2014energy,heilig2015cloud}, with scarce attention paid to field-level approaches and solutions. ``Field-level'' refers to the level of physical devices and functions, and the main goal of this level is ``to provide interactions between the cyberspace (cloud) and the physical world (real field devices)'' \cite{givehchi2013cloud}. In particular, comprehensive theoretical study on field-level cloud-enabled/assisted control strategies is largely missing. 

One notable exception is the cloud-assisted model predictive control (MPC) approach proposed in \cite{skarin2019cloud} and \cite{skarin2020cloud}. In this approach, a constrained linear-quadratic MPC problem is solved in the cloud to generate constraint-admissible control at each sample time instant. An unconstrained linear-quadratic regulator is used as a backup local controller to ensure uninterrupted stabilization of the plant when connectivity is lost or the cloud MPC control is not returned within a prescribed deadline. The work of \cite{skarin2019cloud} and \cite{skarin2020cloud} has focused on practical strategies to handle connectivity loss and cloud service latency. However, control performance and constraint enforcement in the presence of plant-model mismatch as well as connectivity, latency or feasibility issues are not addressed.

Note that networked control system (NCS) is a related but different concept from cloud-assisted control. An NCS is a spatially distributed system for which the communication between sensors, actuators, and controllers is supported by a (typically wireless) network \cite{hespanha2007survey,gupta2009networked}.  Although NCS and cloud-assisted control both use wireless networks for communication and data exchange, one distinguishing feature of cloud-assisted control is that cloud computing is used as the remote, sole controller in NCS while cloud-assisted control aims at integrating both cloud and local computations for enhanced performance.

In this paper, we propose a novel cloud-assisted nonlinear MPC approach for finite-duration control tasks, which are frequently encountered in the automotive and aerospace fields (e.g., autonomous vehicle lane change \cite{hatipoglu2003automated}, spacecraft orbital transfer and landing \cite{haberkorn2004low,acikmese2007convex}, etc). Compared to previous work, our approach is distinguished by the following features. First, we investigate a systematic fusion of cloud-computed control and locally-computed control, where the cloud is used as a value-added service to improve control performance when connectivity is available. Second, in our approach, the system requests a service from the cloud only at the initial time of the control task, instead of requesting cloud service at every sample time instant over the control task duration as in the approaches considered in \cite{heilig2015cloud,skarin2019cloud,skarin2020cloud}. This strategy is motivated by the widely used pay-per-use model of cloud computing services \cite{grossman2009case}, i.e., the user pays for every cloud request and thereby it is desirable to minimize cloud request times. This strategy also implies that our approach does not rely on a low-latency connection to the cloud over the entire control task duration, i.e., our approach requires a lower connection quality and thereby has a wider range of applications. Third, our approach considers a general nonlinear system, with no restrictions to linear models and quadratic cost functions that is considered in \cite{skarin2019cloud,skarin2020cloud}. Furthermore, our approach explicitly addresses model prediction errors due to plant-model mismatch and disturbances, including constraint enforcement in the presence of such prediction errors.

With the aforementioned distinguishing features, the contributions of this paper include:
\begin{enumerate}
    \item We propose a unified framework for cloud-assisted MPC design (see Fig.~\ref{fig:architecture}), with a focus on finite-duration control tasks. The proposed framework systematically integrates cloud and local controls to achieve improved  performance.
        \item We rigorously analyze the feasibility and robust constraint satisfaction for the considered paradigm in the presence of disturbances and cloud/local model prediction errors.
        \item We develop a switching-based fusion policy to systematically combine cloud-computed control trajectory with local shrinking-horizon MPC solutions to minimize the worst-case cost-to-go for enhanced performance.
    \item We verify our theoretical results and demonstrate the effectiveness of our cloud-assisted MPC approach in terms of improving control performance with multiple simulation examples, including an automotive control example to illustrate potential practical application of our approach.
\end{enumerate}

The remainder of this paper is organized as follows. Section~\ref{sec:sys_model} introduces the types of systems and control tasks to be treated as well as the models used by the cloud and local devices to compute controls. We describe the cloud MPC and local MPC designs and analyze their properties in Section~\ref{sec:cloud_local}. We then develop a switching-based fusion scheme to systematically combine cloud and local MPC controls for enhanced performance in Section~\ref{sec:fusion}. We use multiple simulation examples to verify the theoretical results and demonstrate the effectiveness in terms of improving overall control performance of our cloud-assisted MPC approach in Section~\ref{sec:examples}. Conclusions and future work are discussed in Section~\ref{sec:conclusion}.

The notations used in this paper are standard. In particular, we use $\|\cdot\|$ to denote an arbitrary vector norm and its induced matrix norm, and use $\|\cdot\|_Q$ with a positive-definite symmetric matrix $Q$ to denote a quadratic norm, i.e., $\|\cdot\|_Q = \sqrt{(\cdot)^{\top} Q (\cdot)}$. The symbols $\oplus$ and $\sim$ denote, respectively, the Minkowski sum operation and the Minkowski/Pontryagin difference operation between sets \cite{kolmanovsky1998theory}.

\section{System and Models}\label{sec:sys_model}

\begin{figure}[!htbp]
	\centering
		\includegraphics[width=3.4in]{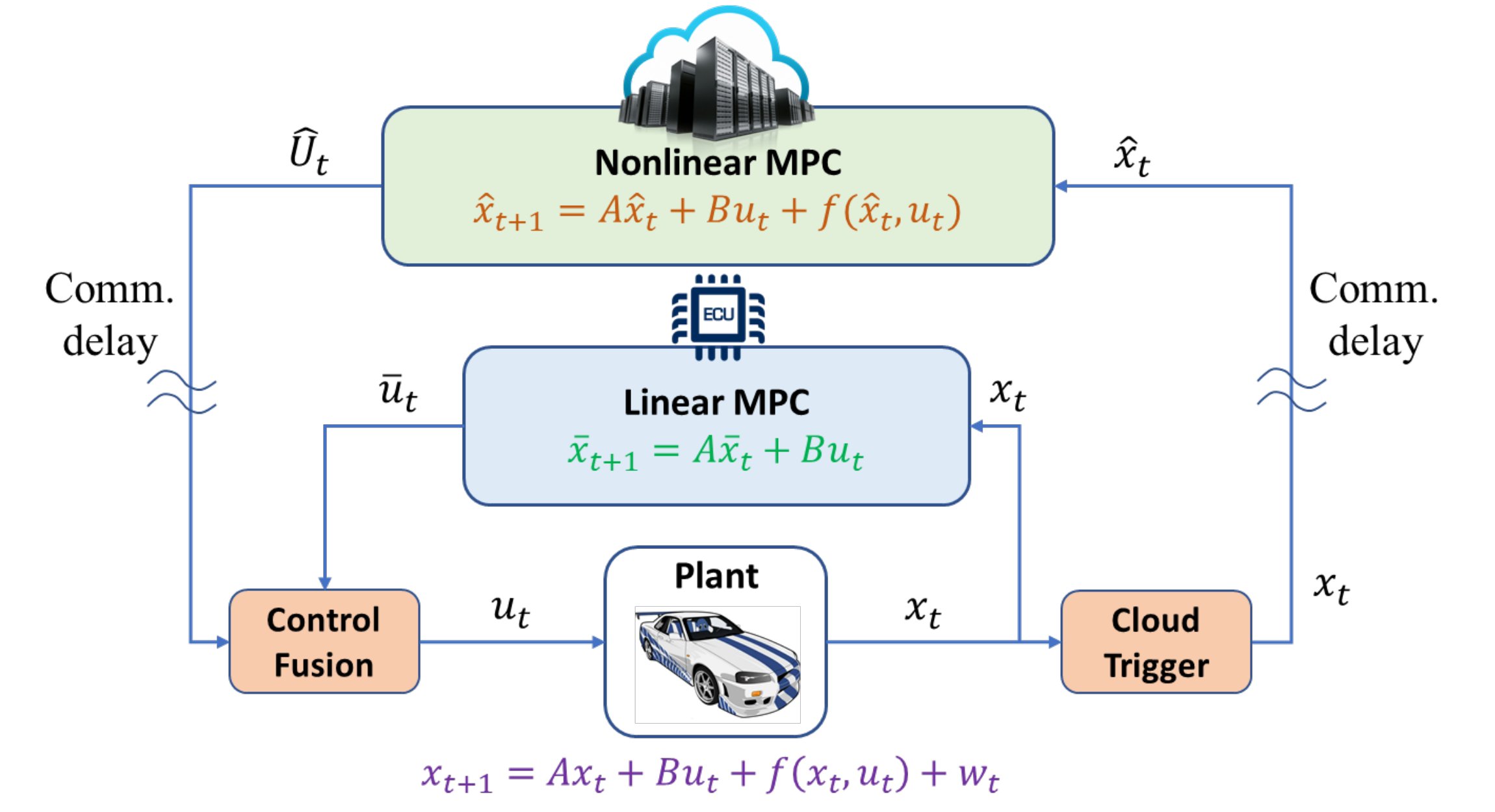}
	\caption{Illustration of the architecture of the proposed  cloud-assisted control framework based on MPC.}
	\label{fig:architecture}
\end{figure}

In this paper, we investigate a new cloud-assisted control paradigm that is illustrated in Fig.~\ref{fig:architecture}. Unlike conventional networked control systems that exploit only one (remote) controller, the considered paradigm seamlessly integrates two complementary computing platforms (cloud and onboard computation units). Specifically, the MPC optimizations are performed both on the cloud and on the local processor (e.g., automotive electronic control unit). On the cloud side, the cloud can support an MPC with a high-fidelity model but is subject to communication delays. On the local agent side, the onboard controller runs an MPC with a simplified model due to limited onboard computations but with negligible time delays. In this work, the two controllers are systematically designed and integrated for enhanced performance.
 
More specifically, we consider a plant that can be represented by the following discrete-time state-space model,
\begin{equation}\label{eqn:sys_1}
    x_{t+1} = A x_t + B u_t + f(x_t,u_t) + w_t,
\end{equation}
where $x_t \in \mathbb{R}^n$ denotes the state of the system at the discrete time instant $t \in \mathbb{N}_0$, $u_t \in \mathbb{R}^m$ denotes the control input at $t$, $A$ and $B$ are matrices of appropriate dimensions, $f: \mathbb{R}^n \times \mathbb{R}^m \to \mathbb{R}^n$ is a nonlinear function, and $w_t \in \mathbb{R}^p$ represents unmeasured disturbances acting on the system.

We make the following assumptions about system \eqref{eqn:sys_1}:

{\bf A1.}  $f$ is a Lipschitz continuous function such that $\|f(x,u) - f(x',u')\| \le L_f \|x-x'\| + M_f \|u-u'\|$ for all $x, x' \in \mathbb{R}^n$ and $u, u' \in \mathbb{R}^m$, where $L_f, M_f \ge 0$ are known Lipschitz constants of $f$; and  $f(0,0) = 0$.

Note that the second condition of {\bf A1}, $f(0,0) = 0$, is equivalent to saying that $(x,u) = (0,0)$ is a nominal steady-state pair and the matrices $A, B$ correspond to a linear model of the system around the steady-state pair $(x,u) = (0,0)$.

{\bf A2.} The disturbance input $w_t$ takes values in a known bounded set $W \subset \mathbb{R}^p$ and $\sup_{w \in W} \|w\| = \omega$.

The control objective is to minimize the following cost function,
\begin{equation}\label{eqn:cost_1}
    J = \sum_{t = 0}^{N-1} \phi(x_t,u_t) + \psi(x_N),
\end{equation}
where $N$ is the optimization horizon that corresponds to the end time of a finite-duration control task (i.e., $t = 0,1,\dots,N$ corresponds to the control task duration). The following assumption  is made about the cost function \eqref{eqn:cost_1}:

{\bf A3.} $\phi(\cdot,\cdot)$ and $\psi(\cdot)$ are both Lipschitz continuous in $x$, i.e., satisfy $|\phi(x,u) - \phi(x',u)| \le L_\phi \|x-x'\|$ and $|\psi(x) - \psi(x')| \le L_\psi \|x-x'\|$ for all $x, x' \in \mathbb{R}^n$, where $L_\phi, L_\psi \ge 0$ are known Lipschitz constants of $\phi$ and $\psi$, respectively.

Being globally Lipschitz is in general a strong assumption on the functions $\phi$ and $\psi$. For example, the commonly used quadratic cost function is not globally Lipschitz. However, if the system state is known to stay within or needs to be constrained within a bounded domain during the system operation, then the global Lipschitz conditions in {\bf A3} can be replaced with local Lipschitz conditions for later developments, which are easier to satisfy. For instance, every continuously differentiable function is Lipschitz continuous on every compact set \cite{sohrab2003basic}.

Meanwhile, constraints that represent hard operational specifications are ubiquitous. Two types of constraints are most common: pointwise-in-time constraints, i.e., constraints that must be satisfied for all time $t = 0,\dots,N$, and terminal constraints, i.e., constraints that must be satisfied at the terminal time $t = N$. For instance, pointwise-in-time constraints are frequently used to represent safety-related requirements, such as variable bounds in chemical processes and collision avoidance in autonomous driving, while terminal constraints are frequently used to represent terminal control objectives in the case of finite-duration tasks and used for stability reasons. In this paper, we consider state constraints taking the following form:
\begin{equation}\label{eqn:constraint_1}
    x_T \in \mathcal{X}_T = \left\{x \in \mathbb{R}^n: G_{T,j}^{\top}\, x \le g_{T,j},\, j = 1, \dots, p_T \right\},
\end{equation}
where $T$ takes values in $1,2,\dots,N$, $G_{T,j} \in \mathbb{R}^n$, $g_{T,j} \in \mathbb{R}$, and $p_T$ is the total number of linear inequalities defining the constraint set $X_T$. Note that since $T$ can take any values in $1,2,\dots,N$, \eqref{eqn:constraint_1} can represent both pointwise-in-time and terminal state constraints.

As shown in Fig.~\ref{fig:architecture}, the cloud MPC exploits the following higher-fidelity model to compute control for the system,
\begin{equation}\label{eqn:model_1}
   \Sigma^c:\,\, \hat{x}_{t+1} = A \hat{x}_t + B u_t + f(\hat{x}_t,u_t).
\end{equation}
Here, we use the ``hat'' notation to represent predicted states in the cloud MPC. This model includes the nonlinear term $f(\hat{x}_t,u_t)$ and necessarily leads to a nonlinear and non-convex optimization problem, which can be computationally intensive. Nevertheless, due to access to powerful cloud computations, we assume this can be solved efficiently on the cloud.

On the other side, due to limited onboard resources, the following lower-fidelity model is used by the local MPC to compute control for the system,
\begin{equation}\label{eqn:model_2}
   \Sigma^l:\,\, \bar{x}_{t+1} = A \bar{x}_t + B u_t,
\end{equation}
which is a linear model and thereby yields easier optimization problems that can be handled by onboard computing devices. We use the ``bar'' notation to represent predicted states in the local MPC.

Our cloud-assisted nonlinear MPC approach is featured by a cloud MPC problem solved at the initial time of the control task, a sequence of shrinking-horizon local MPC problems solved at each sample time instant over the control task duration, and a fusion strategy to systematically combine cloud and local MPC solutions. These components are introduced and analyzed sequentially in the following two sections.

\section{Cloud and Local MPC Designs}\label{sec:cloud_local}

In this section, we introduce the cloud MPC and the local MPC designs and analyze their properties.

\subsection{Cloud MPC Design}\label{sec:cloud_MPC_1}

\begin{figure}[!htbp]
	\centering
		\includegraphics[width=3.4in]{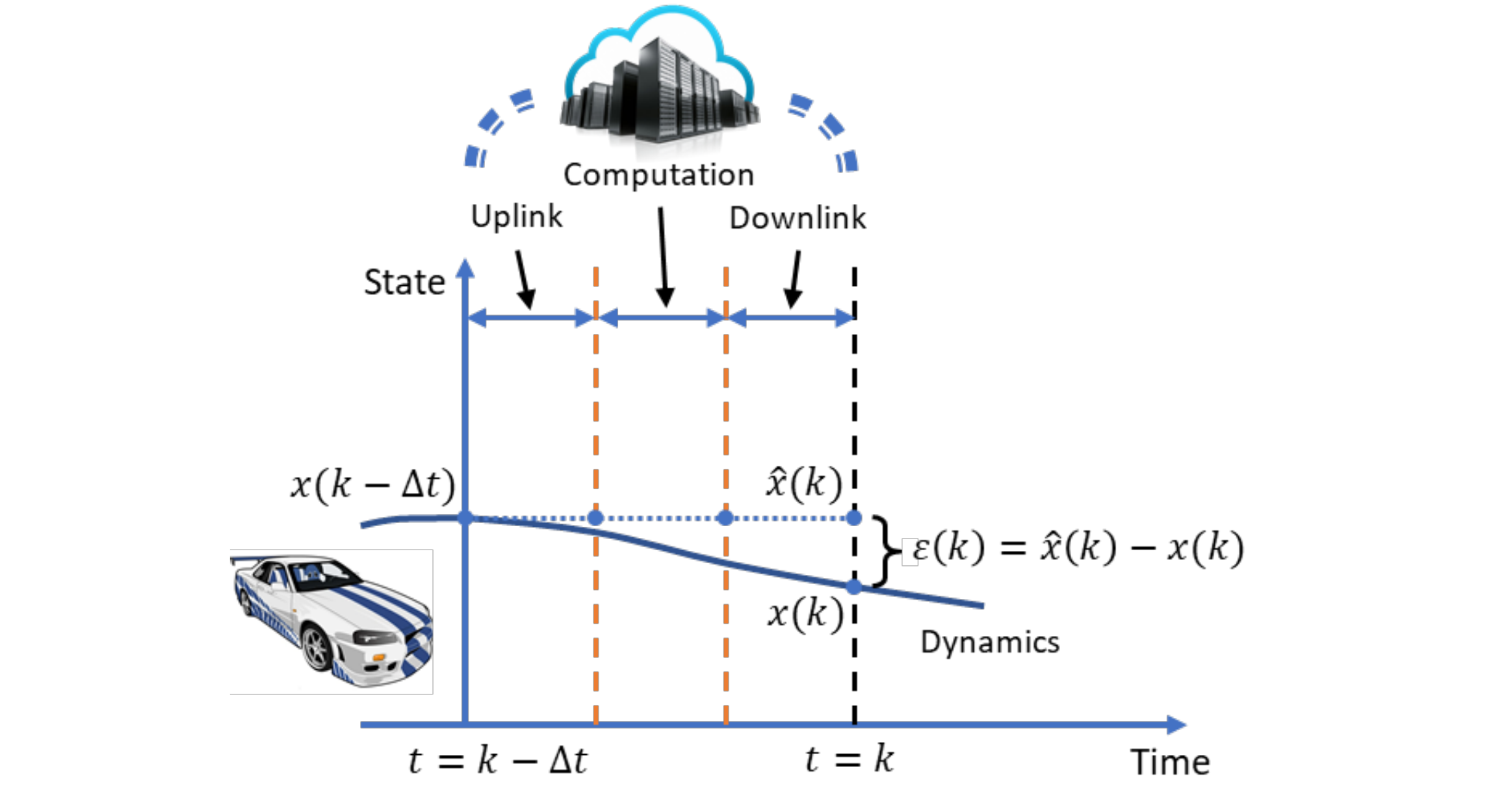}
	\caption{Illustration of cloud state propagation error due to request-response delays.}
	\label{fig:delay}
\end{figure}

When a control task is assigned, we assume that the local system can request the cloud to solve a nonlinear MPC problem based on the higher-fidelity model \eqref{eqn:model_1}. Note that this process incurs a ``request-response delay,'' i.e., there will be a time difference between the local system submitting its computation request to the cloud and receiving the computation results from the cloud, as illustrated in Fig.~\ref{fig:delay}. In the sequel, we refer to such delays as communication delays. While there exist different ways to handle the incurred delays, in this paper, we adopt a ``prediction-ahead-of-time'' approach. Specifically, let $\Delta t$ denote the maximum delay due to cloud communication. Upon cloud MPC request at time instant $t = k - \Delta t$, the cloud will predict the state $\Delta t$ time ahead, $\hat{x}((k - \Delta t) + \Delta t) = \hat{x}(k)$, by assuming state remaining constant over the time interval $\Delta t$ or by running forward simulation based on system model \eqref{eqn:model_1}. The predicted state $\hat{x}(k)$ is then used as the initial condition for the cloud MPC and the obtained optimal control trajectory is downloaded to the local system for use starting at time instant $t = (k - \Delta t) + \Delta t = k$ to avoid any outdated control inputs. This delay treatment strategy essentially converts the effect of communication delays to initial state uncertainty, which is illustrated in Fig.~\ref{fig:delay}.

Without loss of generality, we consider that at time instant $t = - \Delta t$, the local system requests the cloud to perform a nonlinear MPC based on the cost function \eqref{eqn:cost_1}, the model \eqref{eqn:model_1}, and an estimate of state at $t = 0$, $\hat{x}_0$. For now, we assume generic constraints on the predicted states $\hat{x}_\tau$ and controls $u_\tau$. In particular, the cloud is requested to solve the following optimization problem,
\begin{subequations}\label{eqn:cloud_mpc}
\begin{align}
\min\quad & J = \sum_{\tau = 0}^{N-1} \phi(\hat{x}_{\tau},u_{\tau}) + \psi(\hat{x}_N) \\
\text{s.t.}\quad & \hat{x}_{t+1} = A \hat{x}_t + B u_t + f(\hat{x}_t,u_t), \\[4pt]
& \left(\{\hat{x}_{\tau}\}_{\tau = 1}^{N},\{u_{\tau}\}_{\tau = 0}^{N-1}\right) \in \Xi^c, \label{eqn:cloud_mpc_constraint}
\end{align}
\end{subequations}
with the initial condition $\hat{x}_0$. Note that \eqref{eqn:cloud_mpc_constraint} can represent any constraints on $\hat{x}_\tau$ and $u_\tau$. We will elaborate \eqref{eqn:cloud_mpc_constraint} later on to enforce specific state constraints in the form of \eqref{eqn:constraint_1} in a robust manner (i.e., ensuring satisfaction of \eqref{eqn:constraint_1} by the actual system in spite of model prediction errors).

By solving \eqref{eqn:cloud_mpc}, the cloud computes and transmits to the local system an optimal control trajectory available starting from time $0$, $\{\hat{u}_0,\hat{u}_1,\dots,\hat{u}_{N-1}\}$, an associated prediction of state trajectory, $\{\hat{x}_0,\hat{x}_1,\dots,\hat{x}_{N}\}$, and a corresponding sequence of cost-to-go, $\{\hat{J}_0,\hat{J}_1,\dots,\hat{J}_N\}$, where
\begin{equation}
\hat{J}_k = \sum_{\tau = k}^{N-1} \phi(\hat{x}_\tau,\hat{u}_\tau) + \psi(\hat{x}_N).
\end{equation}

Note that if the controls $\{\hat{u}_0,\hat{u}_1,\dots,\hat{u}_{N-1}\}$ are applied to the actual system \eqref{eqn:sys_1}, the resulting actual state trajectory $\{x_0,x_1,\dots,x_{N}\}$ will be different from the predicted trajectory $\{\hat{x}_0,\hat{x}_1,\dots,\hat{x}_{N}\}$, due to the disturbances $w_t$ acting on the actual system \eqref{eqn:sys_1} and errors between $\hat{x}_0$ and $x_0$ caused by imperfect prediction (shown in Fig.~\ref{fig:delay}). Correspondingly, the predicted cost-to-go $\hat{J}_k$, for $k = 0,\dots,N$, will have some errors from the actual cumulative cost over the steps $\tau = k, \dots, N$, which is defined as follows:
\begin{equation}
J_k^c = \sum_{\tau = k}^{N-1} \phi(x_\tau,\hat{u}_\tau) + \psi(x_N),
\end{equation}
where $x_\tau$ denotes the actual state at time $\tau$ under the control input sequence $\{\hat{u}_0,\hat{u}_1,\dots,\hat{u}_{\tau-1}\}$.

The following proposition establishes bounds on the errors between $\hat{x}_\tau$ and $x_\tau$ and between $\hat{J}_k$ and $J_k^c$, which will be exploited later to develop an approach for robust constraint enhancement and to design our cloud-local MPC fusion scheme.

\smallskip
\noindent
{\bf Proposition 1.} Suppose at time $k$, $k \in \{0,1,\dots,N-1\}$, the difference between the predicted state $\hat{x}_k$ and the actual state $x_k$ is $\varepsilon_k = \hat{x}_k - x_k$ and the control sequence $\{\hat{u}_k,\dots,\hat{u}_{N-1}\}$ is applied to the actual system \eqref{eqn:sys_1} over all future steps $\tau = k, \dots, N-1$. Then, the error between the predicted state $\hat{x}_\tau$ and the actual state $x_\tau$, for $\tau = k+1,\dots,N$, can be bounded as
\begin{equation}\label{eqn:P11}
    \|\hat{x}_\tau - x_\tau\| \le (a + L_f)^{\tau-k} \epsilon_k + \sum_{l = k}^{\tau-1} (a + L_f)^{\tau-l-1} \omega,
\end{equation}
where $a = \|A\|$, $\epsilon_k = \|\varepsilon_k\|$, and $\omega = \sup_{w \in W} \|w\|$ (see {\bf A2}). Meanwhile, the error between the predicted cost-to-go $\hat{J}_k$ and the actual cumulative cost $J_k^c$ can be bounded as
\begin{align}\label{eqn:P12}
    & |\hat{J}_k - J_k^c| \le \\
    & \frac{M_k \big((a + L_f)^{N-k} - 1\big) \!-\! (N-k) L_\phi \omega}{(a + L_f) - 1} \!+\!  L_\psi \epsilon_k (a + L_f)^{N-k}, \nonumber 
\end{align}
where
\begin{equation}\label{eqn:P13}
   M_k = L_\phi \epsilon_k + L_\psi w + \frac{L_\phi w}{(a + L_f) - 1}.
\end{equation}

\begin{proof} 
Firstly, using \eqref{eqn:sys_1} and \eqref{eqn:model_1} we obtain
\begin{align}\label{eqn:P14}
& \|\hat{x}_{k+1} - x_{k+1}\| = \nonumber \\
& \|A \hat{x}_k + B \hat{u}_k + f(\hat{x}_k,\hat{u}_k) - A x_k - B \hat{u}_k - f(x_k,\hat{u}_k) - w_k\| \nonumber \\
&\le \|A (\hat{x}_k -x_k)\|  + \|f(\hat{x}_k,\hat{u}_k) - f(x_k,\hat{u}_k)\| + \|w_k\| \nonumber \\
&\le (\|A\| + L_f) \|\hat{x}_k -x_k\| + \|w_k\| \nonumber \\ 
&\le (a + L_f) \epsilon_k + \omega,
\end{align}
where we have used the Lipschitz continuity assumption of $f$ (i.e., {\bf A1}) to derive the inequality in the fourth line. This proves \eqref{eqn:P11} for $\tau = k+1$. We now assume \eqref{eqn:P11} holds for $\tau = \sigma$. In this case, following \eqref{eqn:P14} we can obtain
\begin{align}\label{eqn:P15}
& \|\hat{x}_{\sigma+1} - x_{\sigma+1}\| \le (a + L_f) \|\hat{x}_{\sigma} -x_{\sigma}\| + \omega \nonumber \\
&\le (a + L_f) \left((a + L_f)^{\sigma-k} \epsilon_k + \sum_{l = k}^{\sigma-1} (a + L_f)^{\sigma-l-1} \omega \right) + \omega \nonumber \\
&= (a + L_f)^{\sigma+1-k} \epsilon_k + \sum_{l = k}^{\sigma-1} (a + L_f)^{\sigma-l} \omega + \omega \nonumber \\
&= (a + L_f)^{\sigma+1-k} \epsilon_k + \sum_{l = k}^{\sigma} (a + L_f)^{\sigma-l} \omega,
\end{align}
where we have used the inequality \eqref{eqn:P11} for $\tau = \sigma$ to derive the second line. This proves \eqref{eqn:P11} for $\tau = \sigma+1$. Combining the base case in \eqref{eqn:P14} and the induction step in \eqref{eqn:P15}, \eqref{eqn:P11} is proved for all $\tau = k+1,\dots,N$ by induction.

We now estimate the error between $\hat{J}_k$ and $J_k^c$:
\begin{align}\label{eqn:P16}
& |\hat{J}_k \!-\! J_k^c| \le \sum_{\tau = k}^{N-1} |\phi(\hat{x}_\tau,\hat{u}_\tau) \!-\! \phi(x_\tau,\hat{u}_\tau)| \!+\! |\psi(\hat{x}_N) \!-\! \psi(x_N)| \nonumber \\ 
&\le \sum_{\tau = k}^{N-1} L_{\phi} \|\hat{x}_\tau - x_\tau\| + L_{\psi} \|\hat{x}_N - x_N\| \nonumber \\
&\le \sum_{\tau = k}^{N-1} L_{\phi} \left((a + L_f)^{\tau-k} \epsilon_k + \sum_{l = k}^{\tau-1} (a + L_f)^{\tau-l-1} \omega\right) \nonumber \\
&\quad + L_{\psi} \left((a + L_f)^{N-k} \epsilon_k + \sum_{l = k}^{N-1} (a + L_f)^{N-l-1} \omega\right) \nonumber \\
&= L_{\phi} \epsilon_k \frac{(a + L_f)^{N-k}-1}{(a + L_f)-1} + L_{\phi} \omega \sum_{\tau = k}^{N-1} \frac{(a + L_f)^{\tau-k}-1}{(a + L_f)-1} \nonumber \\
&\quad +  L_{\psi} \epsilon_k (a + L_f)^{N-k} + L_{\psi}\omega \frac{(a + L_f)^{N-k}-1}{(a + L_f)-1} \nonumber \\
&= \left(L_{\phi} \epsilon_k + L_{\psi}\omega + \frac{L_{\phi} \omega}{(a + L_f)-1}\right) \frac{(a + L_f)^{N-k}-1}{(a + L_f)-1} \nonumber \\
&\quad - \frac{(N-k)L_{\phi} \omega}{(a + L_f)-1} + L_{\psi} \epsilon_k (a + L_f)^{N-k}.
\end{align}
Therefore, we have shown the bound in \eqref{eqn:P12}.
\end{proof}

Note that the above results hold true for generic constraints on the predicted states $\hat{x}_\tau$ and controls $u_\tau$ imposed in the cloud MPC optimization problem \eqref{eqn:cloud_mpc}. We now consider specific state constraints in the form of \eqref{eqn:constraint_1}. Note that imposing the constraint \eqref{eqn:constraint_1} directly in the cloud MPC problem \eqref{eqn:cloud_mpc} does not guarantee \eqref{eqn:constraint_1} will be satisfied by the actual state, due to prediction errors discussed above. Therefore, in what follows we develop an approach to robustly enforcing \eqref{eqn:constraint_1}. 

We assume that at the initial time $t = 0$, the error between the state estimate $\hat{x}_0$ (i.e., the initial condition of the cloud MPC problem \eqref{eqn:cloud_mpc}) and the actual state $x_0$ can be bounded as
\begin{equation}\label{eqn:cloud_bound_1}
 \|\hat{x}_0 - x_0\| \le \delta_0,
\end{equation}
where $\delta_0$ is a known constant.

Note that when the system state is assumed to be locally measured, the error between $\hat{x}_0$ and $x_0$ is mainly caused by the delay due to cloud communication, which includes uplink, downlink, and computational delays, as illustrated in Fig.~\ref{fig:delay}. In this case, it is possible for the cloud to estimate a bound for $\|\hat{x}_0 - x_0\|$ according to maximum delay length as well as system dynamics and disturbance characteristics.

Under the assumption \eqref{eqn:cloud_bound_1}, we define
\begin{equation}\label{eqn:cloud_bound_2}
    \delta_k = (a + L_f)^k \delta_0 + \sum_{l = 0}^{k-1} (a + L_f)^{k-l-1} \omega,
\end{equation}
for $k = 1,2,\dots,T$, and impose the following constraint in the cloud MPC problem as \eqref{eqn:cloud_mpc_constraint}:
\begin{equation}\label{eqn:constraint_2}
\hat{x}_T \in \mathcal{X}_T \sim \mathcal{B}_{\delta_T},
\end{equation}
where $\sim$ denotes the Minkowski/Pontryagin difference operation \cite{kolmanovsky1998theory}, and $\mathcal{B}_{\delta_T} = \left\{x \in \mathbb{R}^n: \|x\| \le \delta_T\right\}$.

By tightening the constraint according to \eqref{eqn:constraint_2}, we can guarantee the original constraint \eqref{eqn:constraint_1} to be satisfied by the actual state $x_T$. This result is formalized in Lemma~1 in Section~\ref{sec:fusion}. Furthermore, suppose the support function for unit ball $\mathcal{B} = \left\{x \in \mathbb{R}^n: \|x\| \le 1\right\}$, $h_\mathcal{B}: \mathbb{R}^n \to \mathbb{R}$, is available, the constraint \eqref{eqn:constraint_2} can be expressed as
\begin{equation}\label{eqn:constraint_3}
G_{T,j}^{\top}\, \hat{x}_T \le g_{T,j} - \delta_T\, h_\mathcal{B}(G_{T,j}),\quad j = 1, \dots, p_T,
\end{equation}
which are linear inequality constraints on the predicted state~$\hat{x}_T$.

\subsection{Local MPC Design}\label{sec:local_MPC_1}

At each time instant $t = 0,1\dots,N-1$, the plant computes locally an optimal control based on the cost function \eqref{eqn:cost_1}, the model \eqref{eqn:model_2}, and a measurement of current state $x_t$. As before, we start with considering the case with generic constraints on the predicted states $\bar{x}_{\tau|t}$ and controls $u_{\tau|t}$. In this case, the following shrinking-horizon local MPC problem is solved at each $t$ to compute control,
\begin{subequations}\label{eqn:local_mpc}
\begin{align}
\min\quad & J_t = \sum_{\tau = t}^{N-1} \phi(\bar{x}_{\tau|t},u_{\tau|t}) + \psi(\bar{x}_{N|t}) \\
\text{s.t.}\quad & \bar{x}_{\tau+1|t} = A \bar{x}_{\tau|t} + B u_{\tau|t}, \\
& \quad\, \bar{x}_{t|t} = x_t, \\
&\!\!\!\! \left(\{\bar{x}_{\tau|t}\}_{\tau = 1}^{N},\{u_{\tau|t}\}_{\tau = 0}^{N-1}\right) \in \Xi^l_t. \label{eqn:local_mpc_constraint}
\end{align}
\end{subequations}
In the above expressions, we use $(\cdot)_{\tau|t}$ to denote a predicted value of the variable $(\cdot)_\tau$ with the prediction made at the time instant $t$.\footnote{We did not use such notations for cloud MPC variables because the cloud MPC problem is solved only once at the initial time $t = 0$.} The constraint \eqref{eqn:local_mpc_constraint} can represent any constraints on $\bar{x}_{\tau|t}$ and $u_{\tau|t}$ and will be elaborated to robustly enforce specific state constraints in the form of \eqref{eqn:constraint_1} later on. By solving \eqref{eqn:local_mpc}, an optimal control trajectory, $\{\bar{u}_{t|t},\bar{u}_{t+1|t},\dots,\bar{u}_{N-1|t}\}$, an associated prediction of state trajectory, $\{\bar{x}_{t|t},\bar{x}_{t+1|t},\dots,\bar{x}_{N|t}\}$, and a corresponding cost value, $\bar{J}_t$, are obtained.

Note that similar to the case of cloud MPC, if the control sequence $\{\bar{u}_{t|t},\bar{u}_{t+1|t},\dots,\bar{u}_{N-1|t}\}$ is applied to the actual system \eqref{eqn:sys_1}, the resulting actual state trajectory $\{x_t,x_{t+1},\dots,x_{N}\}$ will be different from the predicted trajectory $\{\bar{x}_{t|t},\bar{x}_{t+1|t},\dots,\bar{x}_{N|t}\}$, due to plant-model mismatch and the disturbances $w_\tau$ acting on the actual system \eqref{eqn:sys_1}. Correspondingly, the predicted cost value $\bar{J}_t$ will have some error from the actual cumulative cost over the steps $\tau = t, \dots, N$, which is defined as follows:
\begin{equation}
J_t^l = \sum_{\tau = t}^{N-1} \phi(x_\tau,\bar{u}_{\tau|t}) + \psi(x_N),
\end{equation}
where $x_\tau$ denotes the actual state at time $\tau$ under the control input sequence $\{\bar{u}_{t|t},\bar{u}_{t+1|t},\dots,\bar{u}_{\tau-1|t}\}$.

The following proposition establishes bounds on the errors between $\bar{x}_{\tau|t}$ and $x_\tau$ and between $\bar{J}_t$ and $J_t^l$.

\smallskip
\noindent
{\bf Proposition 2.} Suppose the local MPC control sequence $\{\bar{u}_{t|t},\dots,\bar{u}_{N-1|t}\}$ is applied to the actual system \eqref{eqn:sys_1} over the steps $\tau = t, \dots, N-1$. Then, the error between the predicted state $\bar{x}_{\tau|t}$ and the actual state $x_\tau$, for $\tau = t+1,\dots,N$, can be bounded as:
\begin{equation}\label{eqn:P21}
\begin{aligned}
    \|\bar{x}_{\tau|t} \!-\! x_\tau\| &\le \sum_{l = t}^{\tau-1} (a \!+\! L_f)^{\tau-l-1} \!\left(L_f \|\bar{x}_{l|t}\| \!+\! M_f \|\bar{u}_{l|t}\| \!+\! \omega \right),
    \end{aligned}
\end{equation}
where $a = \|A\|$ and $\omega = \sup_{w \in W} \|w\|$ (see {\bf A2}). Meanwhile, the error between the predicted cost value $\bar{J}_t$ and the actual cumulative cost $J_t^l$ can be bounded as:
\begin{align}\label{eqn:P22}
    & |\bar{J}_t - J_t^l| \le \nonumber \\
    & L_\phi \sum_{\tau = t+1}^{N-1} \left( \sum_{l = t}^{\tau-1} (a + L_f)^{\tau-l-1} \left(L_f \|\bar{x}_{l|t}\| + M_f \|\bar{u}_{l|t}\| + \omega \right) \!\right) \nonumber \\
   & + L_\psi \left( \sum_{l = t}^{N-1} (a + L_f)^{\tau-l-1} \left(L_f \|\bar{x}_{l|t}\| + M_f \|\bar{u}_{l|t}\| + \omega \right) \right).
\end{align}

\begin{proof}
Firstly, using \eqref{eqn:sys_1} and \eqref{eqn:model_2} we obtain
\begin{align}\label{eqn:P23}
& \|\bar{x}_{t+1|t} - x_{t+1}\| \nonumber \\
&= \|A \bar{x}_{t|t} + B \bar{u}_{t|t} - A x_t - B \bar{u}_{t|t} - f(x_t,\bar{u}_{t|t}) - w_t\| \nonumber \\
&\le \|f(x_t,\bar{u}_{t|t}) - f(0,0)\| + \|w_t\| \nonumber \\
&\le L_f \|x_t\| + M_f \|\bar{u}_{t|t}\|+ \omega \nonumber \\
&= L_f \|\bar{x}_{t|t}\| + M_f \|\bar{u}_{t|t}\|+ \omega,
\end{align}
where we have used $\bar{x}_{t|t} = x_t$ and $f(0,0) = 0$ (see {\bf A1}) to obtain the third line, used the Lipschitz continuity assumption of $f$ (i.e., {\bf A1}) to obtain the fourth line, and used $\bar{x}_{t|t} = x_t$ again to obtain the last line. This proves \eqref{eqn:P21} for $\tau = t+1$. We now assume \eqref{eqn:P21} holds for $\tau = \sigma$. In this case, following \eqref{eqn:P23} we can obtain
\begin{align}\label{eqn:P24}
& \|\bar{x}_{\sigma+1|t} - x_{\sigma+1}\| \nonumber \\
&= \|A \bar{x}_{\sigma|t} + B \bar{u}_{\sigma|t} - A x_\sigma - B \bar{u}_{\sigma|t} - f(x_\sigma,\bar{u}_{\sigma|t}) - w_\sigma\| \nonumber \\
&\le \|A(\bar{x}_{\sigma|t} - x_\sigma)\| + \|f(x_\sigma,\bar{u}_{\sigma|t}) - f(0,0)\| + \|w_\sigma\| \nonumber \\
&\le a \|\bar{x}_{\sigma|t} - x_\sigma\| + L_f \|x_\sigma - \bar{x}_{\sigma|t} + \bar{x}_{\sigma|t}\| + M_f \|\bar{u}_{\sigma|t})\|+ \omega \nonumber \\
&\le (a+L_f)\|\bar{x}_{\sigma|t} - x_\sigma\| + L_f \|\bar{x}_{\sigma|t}\| + M_f \|\bar{u}_{\sigma|t})\|+ \omega \nonumber \\
&\le (a \!+\! L_f) \left( \sum_{l = t}^{\sigma-1} (a \!+\! L_f)^{\sigma-l-1} \left(L_f \|\bar{x}_{l|t}\| + M_f \|\bar{u}_{l|t}\| \!+\! \omega \right) \!\right) \nonumber \\
&\quad + L_f \|\bar{x}_{\sigma|t}\| + M_f \|\bar{u}_{\sigma|t})\|+ \omega \nonumber \\
&= \sum_{l = t}^{\sigma} (a + L_f)^{\sigma-l-1} \left(L_f \|\bar{x}_{l|t}\| + M_f \|\bar{u}_{l|t}\| + \omega \right),
\end{align}
where we have used the inequality \eqref{eqn:P21} for $\tau = \sigma$ to derive the fifth line. This proves \eqref{eqn:P21} for $\tau = \sigma+1$. Combining the base case in \eqref{eqn:P23} and the induction step in \eqref{eqn:P24}, \eqref{eqn:P21} is proved for all $\tau = t+1,\dots,N$ by induction.

Using the Lipschitz continuity assumptions of $\phi$ and $\psi$ (i.e., {\bf A3}), we can bound the difference between $\bar{J}_t$ and $J_t^l$ according to
\begin{align}
& |\bar{J}_t - J_t^l| = \nonumber \\
    & \left|\sum_{\tau = t}^{N-1} \left(\phi(\bar{x}_{\tau|t},\bar{u}_{\tau|t}) - \phi(x_\tau,\bar{u}_{\tau|t})\right) + \left(\psi(\bar{x}_{N|t}) - \psi(x_N)\right)\right| \nonumber \\
    &\le \sum_{\tau = t}^{N-1} \left|\phi(\bar{x}_{\tau|t},\bar{u}_{\tau|t}) - \phi(x_\tau,\bar{u}_{\tau|t})\right| + \left|\psi(\bar{x}_{N|t}) - \psi(x_N)\right| \nonumber \\
    &\le L_\phi \sum_{\tau = t}^{N-1} \|\bar{x}_{\tau|t} - x_\tau\| + L_\psi \|\bar{x}_{N|t} - x_N\| \nonumber \\
    &= L_\phi \sum_{\tau = t+1}^{N-1} \|\bar{x}_{\tau|t} - x_\tau\| + L_\psi \|\bar{x}_{N|t} - x_N\|,
\end{align}
where we have used $\bar{x}_{t|t} = x_t$ to drop the term for $\tau = t$ in the sum and obtain the last line. Then, using the bounds for $\|\bar{x}_{\tau|t} - x_\tau\|$ in \eqref{eqn:P21} for $\tau = t+1,\dots,N$, we can obtain the bound for $|\bar{J}_t - J_t^l|$ in \eqref{eqn:P22}.
\end{proof}

\smallskip
\noindent
{\bf Remark 1.} The bounds in \eqref{eqn:P21} and \eqref{eqn:P22} depend on the predicted states $\bar{x}_{l|t}$ and controls $\bar{u}_{l|t}$. Note that once the local MPC problem \eqref{eqn:local_mpc} is solved, $\bar{x}_{l|t}$ and $\bar{u}_{l|t}$, for $l = t,\dots,N$, are available. This means the bounds in \eqref{eqn:P21} and \eqref{eqn:P22} are available once \eqref{eqn:local_mpc} is solved.

\smallskip
\noindent
{\bf Remark 2.} Proposition~2 represents a strategy for estimating the actual cost $J_t^l$ corresponding to the control sequence $\{\bar{u}_{t|t},\bar{u}_{t+1|t},\dots,\bar{u}_{N-1|t}\}$ using a cost value predicted by the lower-fidelity linear model \eqref{eqn:model_2}, $\bar{J}_t$, which is obtained along with the control sequence $\{\bar{u}_{t|t},\bar{u}_{t+1|t},\dots,\bar{u}_{N-1|t}\}$ when the local MPC problem \eqref{eqn:local_mpc} is solved. An alternative strategy is to compute another estimate of $J_t^l$ via the higher-fidelity model \eqref{eqn:model_1} by applying the local MPC obtained control sequence $\{\bar{u}_{t|t},\bar{u}_{t+1|t},\dots,\bar{u}_{N-1|t}\}$ to \eqref{eqn:model_1}. Note that this alternative strategy involves an additional step of simulating the higher-fidelity model \eqref{eqn:model_1}, which may take non-negligible effort of local onboard computation units especially for high-dimensional models. Further investigation of this alternative strategy is left to our future work. 

Note that the above results hold true for generic constraints on the predicted states $\bar{x}_{\tau|t}$ and controls $u_{\tau|t}$ imposed in the local MPC optimization problem \eqref{eqn:local_mpc}. We now elaborate \eqref{eqn:local_mpc_constraint} to address specific state constraints in the form of \eqref{eqn:constraint_1}. Similarly as in cloud MPC, in order to guarantee \eqref{eqn:constraint_1} to be satisfied by the actual system \eqref{eqn:sys_1}, robust constraint enforcement techniques for local MPC are needed and are developed in what follows.

To begin with, we consider the following polyhedral approximations of unit ball,
\begin{align}\label{eqn:local_bound}
    \bar{\mathcal{X}} &= \left\{x \in \mathbb{R}^n: \bar{G} x \le \bar{g} \right\}, \nonumber \\
    \bar{\mathcal{U}} &= \left\{u \in \mathbb{R}^m: \bar{H} u \le \bar{h} \right\},
\end{align}
such that $\|x\| \le 1$ for all $x \in \bar{\mathcal{X}}$ and $\|u\| \le 1$ for all $u \in \bar{\mathcal{U}}$.

We then impose the following constraints on the predicted states and controls in the local MPC problem,
\begin{equation}\label{eqn:constraint_4}
    \bar{x}_{\tau|t} \in \alpha_{\tau|t}\, \bar{\mathcal{X}}, \quad \bar{u}_{\tau|t} \in \beta_{\tau|t}\, \bar{\mathcal{U}},
\end{equation}
for $\tau = t, \dots, T-1$, where $\alpha_{\tau|t}$ and $\beta_{\tau|t}$ are auxiliary decision variables (i.e., to be optimized together with the state and control variables in the local MPC problem). The constraints in \eqref{eqn:constraint_4} can be expressed as the following linear inequalities in $(\bar{x}_{\tau|t},\alpha_{\tau|t})$ and $(\bar{u}_{\tau|t},\beta_{\tau|t})$:
\begin{align}\label{eqn:constraint_5}
    \bar{G} \bar{x}_{\tau|t} - \alpha_{\tau|t} \bar{g} \le 0, \nonumber \\
    \bar{H} \bar{u}_{\tau|t} - \beta_{\tau|t} \bar{h} \le 0.
\end{align}

Under \eqref{eqn:constraint_4} or \eqref{eqn:constraint_5}, we have $\|\bar{x}_{\tau|t}\| \le \alpha_{\tau|t}$ and $\|\bar{u}_{\tau|t}\| \le \beta_{\tau|t}$ for $\tau = t, \dots, T-1$. Then, using \eqref{eqn:P21}, the difference between the predicted state $\bar{x}_{T|t}$ and the actual state $x_T$ can be bounded as
\begin{equation}\label{eqn:local_bound_1}
    \|\bar{x}_{T|t} - x_T\| \le \sum_{l = t}^{T-1} (a + L_f)^{T-l-1} \left(L_f \alpha_{l|t} + M_f \beta_{l|t} + \omega \right).
\end{equation}

To ensure the constraint \eqref{eqn:constraint_1} will be satisfied by the actual state $x_T$, we finally impose the following constraint in the local MPC problem:
\begin{equation}\label{eqn:constraint_6}
\bar{x}_{T|t} \in \mathcal{X}_T \sim \mathcal{B}_{\xi_{T|t}},
\end{equation}
where $\mathcal{B}_{\xi_{T|t}} = \left\{x \in \mathbb{R}^n: \|x\| \le \xi_{T|t}\right\}$ and $\xi_{T|t}$ denotes the bound for $\|\bar{x}_{T|t} - x_T\|$ on the right-hand side of \eqref{eqn:local_bound_1}. 

Similar to \eqref{eqn:constraint_3}, the constraint \eqref{eqn:constraint_6} can be expressed as the following linear inequalities in $\big(\bar{x}_{T|t},\{\alpha_{\tau|t},\beta_{\tau|t}\}_{\tau = t,\dots,T-1}\big)$ using the support function for unit ball $h_{\mathcal{B}}$:
\begin{align}\label{eqn:constraint_7}
& G_{T,j}^{\top}\, \bar{x}_{T|t} + h_\mathcal{B}(G_{T,j})\, \xi_{T|t} = G_{T,j}^{\top}\, \bar{x}_{T|t} \nonumber \\
& + h_\mathcal{B}(G_{T,j}) \left(\,\sum_{l = t}^{T-1} (a + L_f)^{T-l-1} \left(L_f \alpha_{l|t} + M_f \beta_{l|t} + \omega \right)\right) \nonumber \\
&\quad\quad\quad \le g_{T,j}, \quad\quad\quad j = 1, \dots, p_T.
\end{align}

The robust constraint enforcement property of the (state-) constrained local MPC problem (i.e., \eqref{eqn:local_mpc} with the generic constraint \eqref{eqn:local_mpc_constraint} elaborated as \eqref{eqn:constraint_4} and \eqref{eqn:constraint_6}) is further formalized in Lemma~2 in Section~\ref{sec:fusion}. Note that this problem is reformulated according to the current state measurement $x_t$ at every sample time instant $t = 0,1\dots,N-1$. It may not be feasible at all times. If the constrained local MPC problem is infeasible at some time $t$, as a fail-safe solution, we let
\begin{equation}\label{eqn:fail_safe}
\bar{u}_{\tau|t} = \bar{u}_{\tau|t-1},
\end{equation}
for $\tau = t,\dots,N-1$. This fail-safe solution ensures that the local controller produces a control signal for the plant to apply at every time step.

\section{Switching-Based Control Fusion}\label{sec:fusion}

A distinguishing component of our cloud-assisted nonlinear MPC approach is a fusion scheme to systematically combine the cloud and local MPC controls obtained in Sections~\ref{sec:cloud_local} to achieve enhanced performance. We describe our proposed switching-based fusion scheme in this section. We first present a simpler version of our fusion policy for the case without state constraints in Section~\ref{sec:policy_1}, then present a modified version for the case with state constraints in Section~\ref{sec:policy_2}, together with analysis of its feasibility and constraint satisfaction properties.

\subsection{Switch Policy for Unconstrained Case}\label{sec:policy_1}

At each time instant $t = 0,1\dots,N-1$, two candidate controls, $\hat{u}_t$ from cloud MPC and $\bar{u}_{t|t}$ from local MPC, are available. The goal is to systematically fuse them to achieve enhanced performance. The fundamental idea of the proposed switching-based fusion strategy is to minimize future cumulative cost: one should apply $\hat{u}_t$ if $J_t^c \le J_t^l$ and apply $\bar{u}_{t|t}$ otherwise. However, only the estimates of $J_t^c$ and $J_t^l$, i.e., $\hat{J}_t$ and $\bar{J}_t$, but themselves, are available, and these estimates have errors. Therefore, following the idea of minimizing the worst-case cost, we propose the following switching policy for the case without state constraints:
\begin{align}\label{eqn:policy_1}
    u_t = \begin{cases} \hat{u}_t, & \text{if } \hat{J}_t + \hat{\eta}_t \le \bar{J}_t + \bar{\eta}_t, \\ \bar{u}_{t|t}, & \text{otherwise}, \end{cases}
\end{align}
where $\hat{\eta}_t$ denotes the bound for $|\hat{J}_t - J_t^c|$ on the right-hand side of \eqref{eqn:P12} and $\bar{\eta}_t$ denotes the bound for $|\bar{J}_t - J_t^l|$ on the right-hand side of \eqref{eqn:P22}. This switching policy completes the unconstrained version of our proposed cloud-assisted MPC approach to finite-duration control tasks.

The switching policy \eqref{eqn:policy_1} uses the bounds $\hat{\eta}_t$ and $\bar{\eta}_t$ developed in \eqref{eqn:P12} and \eqref{eqn:P22} based on Lipschitz constants of relevant functions to estimate the worst-case costs and determine an appropriate control to apply. The proposed switching strategy is essentially based on the worst-case performance bounds and thus may not be the optimal policy (i.e., the policy that minimizes the actual cumulative cost). Note that computation of optimal policies requires exact evaluation of disturbance effects propagating through the nonlinear system \eqref{eqn:sys_1}, which is in general a difficult task, if not impossible. On the other hand, we show through simulation examples in Section~\ref{sec:examples} that our switching policy \eqref{eqn:policy_1} can lead to improved control performance than using solely cloud MPC or local MPC controls. Investigation into methods to compute/approximate optimal policies is left to our future work.

\subsection{Switch Policy Accounting for Constraints}\label{sec:policy_2}

To guarantee satisfaction of the constraint \eqref{eqn:constraint_1} by the actual state $x_T$, we modify the switch policy \eqref{eqn:policy_1} for $0 \le t \le T-1$ to the following policy:
\begin{align}\label{eqn:policy_2}
    u_t = \begin{cases} \hat{u}_t, & \text{if } \hat{J}_t + \hat{\eta}_t \le \bar{J}_t + \bar{\eta}_t \text{ and } \|\hat{x}_t - x_t\| \le \delta_t, \\ \bar{u}_{t|t}, & \text{otherwise}, \end{cases}
\end{align}
where $\delta_t$ is defined in \eqref{eqn:cloud_bound_2} and we let $\bar{J}_t = + \infty$ if the local MPC problem is infeasible at $t$.

We now discuss the constraint satisfaction property of the policy \eqref{eqn:policy_2}. Our main result, Proposition~3, is built upon two intermediate results, Lemmas~1 and 2, which formalize the robust constraint enforcement properties of the cloud and local MPC problems, respectively.

\smallskip
\noindent
{\bf Lemma 1.} Suppose (i) \eqref{eqn:constraint_2} is enforced in the cloud MPC problem (i.e., when the cloud MPC controls $\{\hat{u}_0,\dots,\hat{u}_{N-1}\}$ are computed), (ii) at time $k$, for some $0 \le k \le T-1$, the difference between the predicted state $\hat{x}_k$ and the actual state $x_k$ is bounded as $\|\hat{x}_k - x_k\| \le \delta_k$, where $\delta_k$ is defined in \eqref{eqn:cloud_bound_2}, and (iii) the cloud MPC controls $\{\hat{u}_k,\dots,\hat{u}_{T-1}\}$ are applied to the actual system \eqref{eqn:sys_1} over the steps $\tau = k, \dots, T-1$. Then, (I) the error between the predicted state $\hat{x}_T$ and the actual state $x_T$ is bounded as $\|\hat{x}_T - x_T\| \le \delta_T$, and (II) the constraint \eqref{eqn:constraint_1} is satisfied by the actual state $x_T$.

\begin{proof}
The proof of (I) follows similar steps as \eqref{eqn:P14} and \eqref{eqn:P15}, with $\epsilon_k$ replaced by $\delta_k$. Then, since \eqref{eqn:constraint_2} is enforced and $\|\hat{x}_T - x_T\| \le \delta_T$, we must have $x_T \in \hat{x}_T \oplus \mathcal{B}_{\delta_T} \subseteq (\mathcal{X}_T \sim \mathcal{B}_{\delta_T}) \oplus \mathcal{B}_{\delta_T} \subseteq \mathcal{X}_T$, where $\oplus$ denotes the Minkowski sum operation \cite{kolmanovsky1998theory}. This proves (II).
\end{proof}

\smallskip
\noindent
{\bf Lemma 2.} Suppose at time $t$, $0 \le t \le T-1$, the constraints \eqref{eqn:constraint_4} and \eqref{eqn:constraint_6} are enforced in the local MPC problem and a sequence of (feasible) optimal controls $\{\bar{u}_{t|t},\dots,\bar{u}_{N-1|t}\}$ are obtained. Then, if $\{\bar{u}_{t|t},\dots,\bar{u}_{T-1|t}\}$ are applied to the actual system \eqref{eqn:sys_1} over the steps $\tau = t, \dots, T-1$, the constraint \eqref{eqn:constraint_1} is necessarily satisfied by the actual state $x_T$.

\begin{proof}
Let us denote the $\alpha_{\tau|t}, \beta_{\tau|t}$ values associated with the solution $\{\bar{u}_{t|t},\dots,\bar{u}_{N-1|t}\}$ as $\bar{\alpha}_{\tau|t}, \bar{\beta}_{\tau|t}$ and let
\begin{equation}
\bar{\xi}_{T|t} = \sum_{l = t}^{T-1} (a + L_f)^{T-l-1} \left(L_f \bar{\alpha}_{l|t} + M_f \bar{\beta}_{l|t} + \omega \right).
\end{equation}
If $\{\bar{u}_{t|t},\dots,\bar{u}_{T-1|t}\}$ are applied to the actual system over the steps $\tau = t, \dots, T-1$, then according to \eqref{eqn:local_bound_1} we  have
\begin{equation}
\|\bar{x}_{T|t} - x_T\| \le \bar{\xi}_{T|t},    
\end{equation}
where $\bar{x}_{T|t}$ denotes the local MPC predicted state corresponding to the controls $\{\bar{u}_{t|t},\dots,\bar{u}_{T-1|t}\}$. Meanwhile, the constraint \eqref{eqn:constraint_6} ensures
\begin{equation}
\bar{x}_{T|t} \in \mathcal{X}_T \sim \mathcal{B}_{\bar{\xi}_{T|t}}.    
\end{equation}
Therefore, we have $x_T \in \bar{x}_{T|t} \oplus \mathcal{B}_{\bar{\xi}_{T|t}} \subseteq (\mathcal{X}_T \sim \mathcal{B}_{\bar{\xi}_{T|t}}) \oplus \mathcal{B}_{\bar{\xi}_{T|t}} \subseteq \mathcal{X}_T$. This proves the result.
\end{proof}

On the basis of the above two results, we now show that the modified switching policy \eqref{eqn:policy_2} leads to the following constraint satisfaction result:

\smallskip
\noindent
{\bf Proposition 3.} Suppose (i) at the initial time $t = 0$, the state-constrained cloud MPC problem (i.e., \eqref{eqn:cloud_mpc} with the generic constraint \eqref{eqn:cloud_mpc_constraint} elaborated as \eqref{eqn:constraint_2}) is feasible, and (ii) at each time instant $t = 0,\dots,T-1$, the control that is applied to the actual system \eqref{eqn:sys_1} is determined by the switching policy \eqref{eqn:policy_2}. Then, the constraint \eqref{eqn:constraint_1} is necessarily satisfied by the actual state~$x_T$.

\begin{proof}
Firstly, following the proof of Lemma~1(I), it can be shown that if $\|\hat{x}_t - x_t\| \le \delta_t$ holds at some $t$, $0 \le t \le T-1$, and $u_t = \hat{u}_t$, then the resulting actual state $x_{t+1}$ must satisfy $\|\hat{x}_{t+1} - x_{t+1}\| \le \delta_{t+1}$. In this case, according to the switching policy \eqref{eqn:policy_2}, if the control is switched from cloud MPC control to local MPC control at some $t$ (i.e., $u_{t-1} = \hat{u}_{t-1}$ and $u_t = \bar{u}_{t|t}$), it must hold that $\bar{J}_t + \bar{\eta}_t < \hat{J}_t + \hat{\eta}_t$, which implies $\bar{J}_t < + \infty$, i.e., the constrained local MPC problem is feasible at $t$.

Let us now consider two cases for the control at time $t = T-1$: (a) If $u_{T-1} = \hat{u}_{T-1}$, according to \eqref{eqn:policy_2}, it must hold that $\|\hat{x}_{T-1} - x_{T-1}\| \le \delta_{T-1}$. In this case, according to Lemma~1(II), we must have $x_T \in \mathcal{X}_T$. (b) If $u_{T-1} = \bar{u}_{T-1|T-1}$, according to the analysis in the first paragraph of this proof, there must exist some $\tau$, $0 \le \tau \le T-1$, such that $\tau$ is the last time instant where the constrained local MPC problem is feasible. Moreover, according to the fail-safe policy \eqref{eqn:fail_safe}, we must have $u_{\tau} = \bar{u}_{\tau|\tau}$, $u_{\tau+1} = \bar{u}_{\tau+1|\tau+1} = \bar{u}_{\tau+1|\tau}$, $\dots$, $u_{T-1} = \bar{u}_{T-1|T-1} = \bar{u}_{T-1|\tau}$. To see this more clearly, if the constrained local MPC problem is feasible at $T-1$, then $\tau = T-1$ and the above statement holds true. In the case where the constrained local MPC problem is infeasible at $T-1$, the analysis in the first paragraph of this proof says that $u_{T-2}$ cannot take the cloud MPC control $\hat{u}_{T-2}$ (because in that case the control would not switch to local MPC control at $t = T-1$ when the local MPC problem is infeasible at $T-1$). Therefore, in this case we must have $u_{T-2} = \bar{u}_{T-2|T-2}$, and, according to the fail-safe policy \eqref{eqn:fail_safe}, $u_{T-1} = \bar{u}_{T-1|T-1} = \bar{u}_{T-1|T-2}$. By continuing this analysis, we can show the statement above. Then, according to Lemma~2, we must have $x_T \in \mathcal{X}_T$. This completes the proof.
\end{proof}

\smallskip
\noindent
{\bf Remark 3.} All of our above developments and theoretical results, including Propositions~1--3, apply to the cases with or without control constraints $u_t \in \mathcal{U}$. Such control constraints can be handled by imposing them directly in the cloud and local MPC optimization problems without extra treatments.

\smallskip
\noindent
{\bf Remark 4 (Closed-Loop Stability).} The robust constraint enforcement approaches developed in Sections~\ref{sec:cloud_local}-A, -B, and \ref{sec:fusion}-B can be used to establish closed-loop stability of the system in continued operation. Specifically, one can design a local controller, $u = \pi(x)$, and find a set, $\mathcal{X}_f$, such that the controller $\pi$ stabilizes the system for states $x$ within the set $\mathcal{X}_f$. Then, one can impose $\mathcal{X}_f$ as a terminal constraint set and apply the robust constraint enforcement approaches developed above to enforce $x_N \in \mathcal{X}_f$. After the system state enters $\mathcal{X}_f$ (which is guaranteed according to Proposition~3), this local controller $\pi$ can be activated to stabilize the system.

\smallskip
\noindent
{\bf Remark 5.} Our approach and associated theoretical results, including the error bounds in Propositions~1 and 2 and the robust constraint satisfaction property in Proposition~3, can be extended and used to treat time-varying systems of the form
\begin{equation}\label{eqn:sys_2}
    x_{t+1} = A_t x_t + B_t u_t + f_t(x_t,u_t) + w_t,
\end{equation}
by defining $a = \max_{t = 0,\dots,N-1} \|A_t\|$, $L_f = \max_{t = 0,\dots,N-1}$ $L_{f,t}$ and $M_f = \max_{t = 0,\dots,N-1} M_{f,t}$ with $L_{f,t}$ and $M_{f,t}$ being the Lipschitz constants of $f_t$. Such a treatment based on over-bounding may lead to conservative control performance and is thus not a focus of this paper. However, this fact can be useful in some applications.

\section{Simulation Examples}\label{sec:examples}

In this section, three simulation examples are presented to demonstrate the effectiveness of our proposed cloud-assisted nonlinear MPC approach.

\subsection{Example 1: First-order system}
 As the first example, we consider a first-order system in the form of \eqref{eqn:sys_1} with the following parameters:
\begin{align}
& A = 0.75, \quad B = 1, \quad f(x_t, u_t) = 0.1x_t - \sin(0.1x_t),
\nonumber \\
& |w_t| \le 0.02, \quad x_0 = -10, \quad \varepsilon_0 = \hat{x}_0 - x_0 = -0.5. 
\end{align}
We consider a control task defined by the following cost function to minimize:
\begin{equation}
J = \sum_{t=0}^{N-1} \left(|x_{t}| + \sqrt{5} |u_{t}|\right) + \sqrt{2} |x_{N}|,
\end{equation}
where $N = 10$ corresponds to the end time of the control task. Note that we consider such a cost function because it is globally Lipschitz, i.e., satisfying our assumption {\bf A3} globally, and thereby facilitates the implementation of our approach and the validation of our theoretical results. Moreover, we assume the control input, $u_t$, and the system state at the terminal time, $x_N$, are subjected to the following constraints:
\begin{align}
& u_t \in \mathcal{U} = \left\{u \in \mathbb{R}: |u| \le 3 \right\}, t = 0, 1, \dots, N-1, \nonumber \\
& x_N \in \mathcal{X}_N = \left\{x \in \mathbb{R}: |x| \le 2.5 \right\}.
\end{align}
The constraint handling techniques presented in Section~\ref{sec:cloud_local} are used to robustly enforce the above constraints. At the initial time $t = 0$, the tightened constraint sets defined in \eqref{eqn:constraint_2} and \eqref{eqn:constraint_6} are $\mathcal{X}_N \sim \mathcal{B}_{\delta_N} = \left\{x \in \mathbb{R}: |x| \le 2.0401 \right\}$ and $\mathcal{X}_N \sim \mathcal{B}_{\xi_{N|0}} = \left\{x \in \mathbb{R}: |x| \le 0.3510 \right\}$, respectively. To clearly demonstrate that our proposed strategy of fusing the cloud- and local MPC controls using the switching policy \eqref{eqn:policy_2} can effectively improve the overall control performance, we also implement ``sole cloud MPC'' and ``sole local MPC'' (i.e., the cloud/local MPC solutions are applied over the entire control task duration without switching) for comparison.

\begin{figure}[!htbp]
	\centering
	\includegraphics[width=2.5 in]{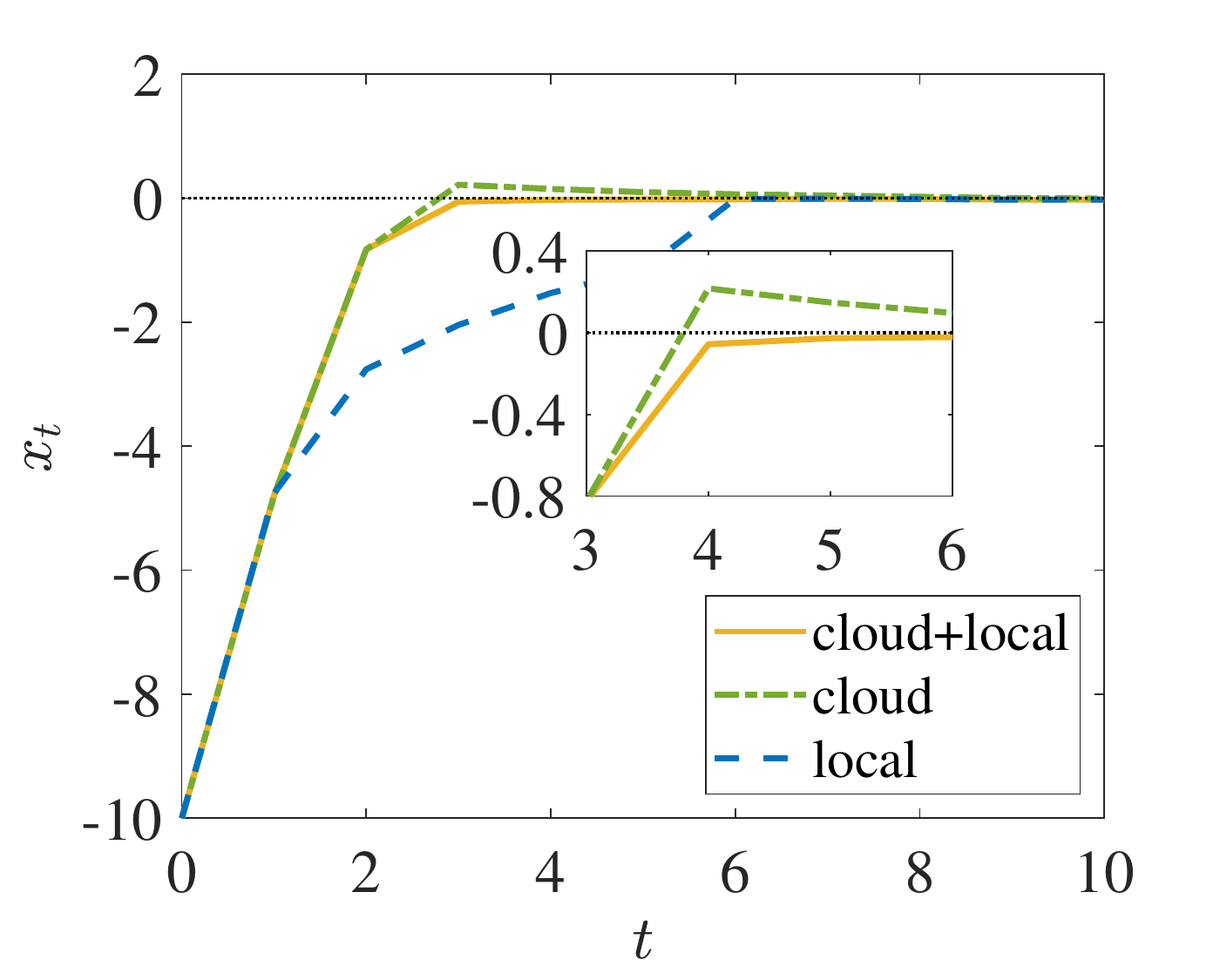}
	\caption{Simulation results of Example 1: State trajectories.}
	\label{fig_simple}
\end{figure}

Fig.~\ref{fig_simple} shows the simulated trajectories of system state $x_t$ corresponding to different MPC schemes. It can be seen that our proposed approach of combining cloud and local MPC controls leads to the best transient response. The mean regulation error ($\text{MRE} = \frac{1}{N+1}\sum_{t = 0}^N |x_t|$) and the resulting actual cost value of each method are summarized in Table~\ref{table MRE}, where it can be seen that our cloud  and local MPC fusion strategy achieves the smallest RMS and cost.

\begin{table}[!t]
	\caption{Mean regulation errors and actual cost values of different MPC approaches}
	\label{table MRE}
	\begin{center}
		\begin{tabular}{c c c c}
			\hline
			\hline
			& cloud + local & cloud & local \\
			\hline
			MRE & 1.5822 & 1.6219 & 2.2344 \\
			\hline
			cost & 29.9165 & 30.7720 & 32.8074 \\
			\hline
			\hline
		\end{tabular}
	\end{center}\vspace{-15pt}
\end{table}

\begin{figure}[!htbp]
	\centering
	\hspace{-0.16 in}
	\subfigure[] {\label{fig_J_simple}
		\includegraphics[width=1.76 in]{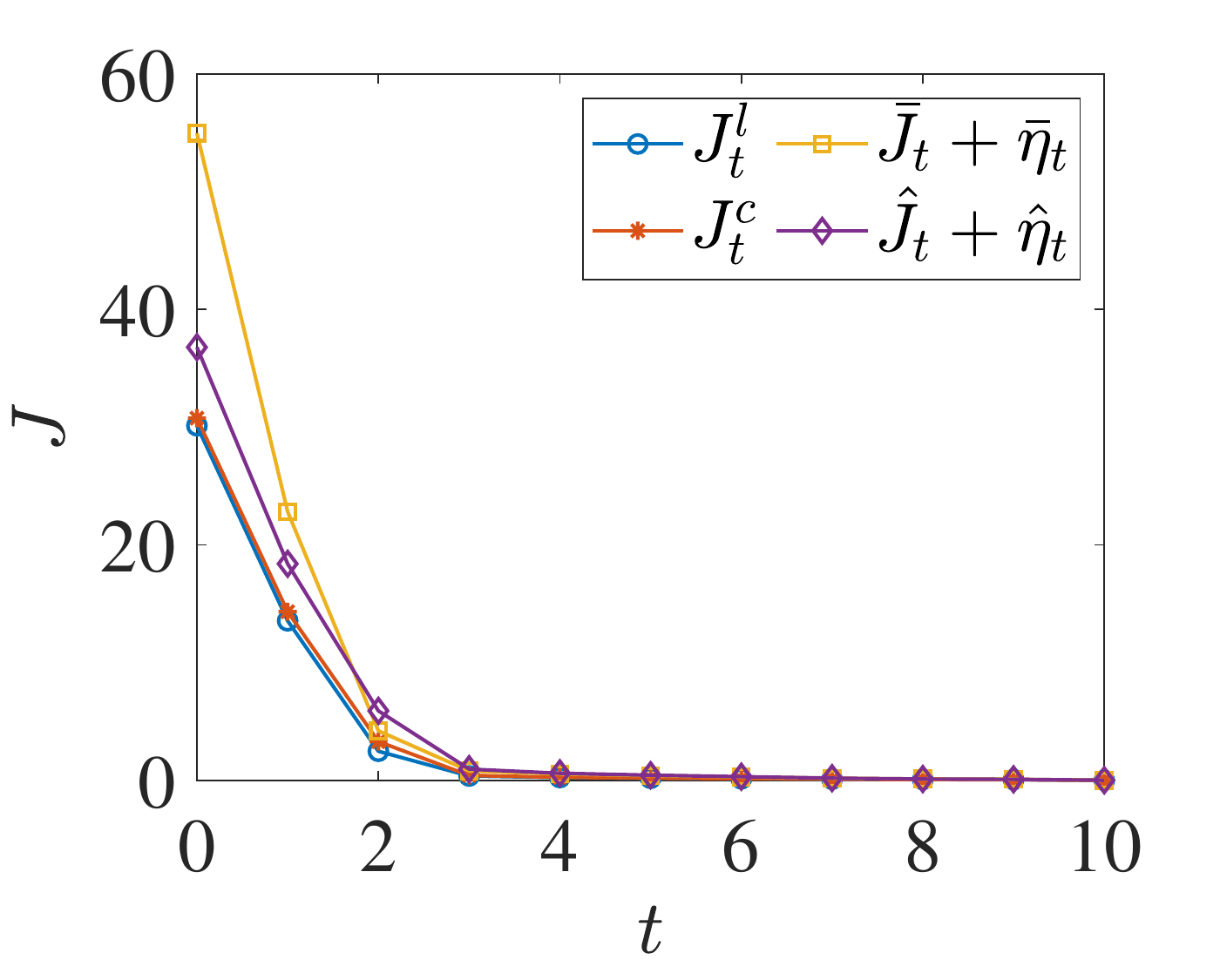}
	}
	\hspace{-0.29 in}
	\subfigure[] {\label{fig_switch_simple}
		\includegraphics[width=1.76 in]{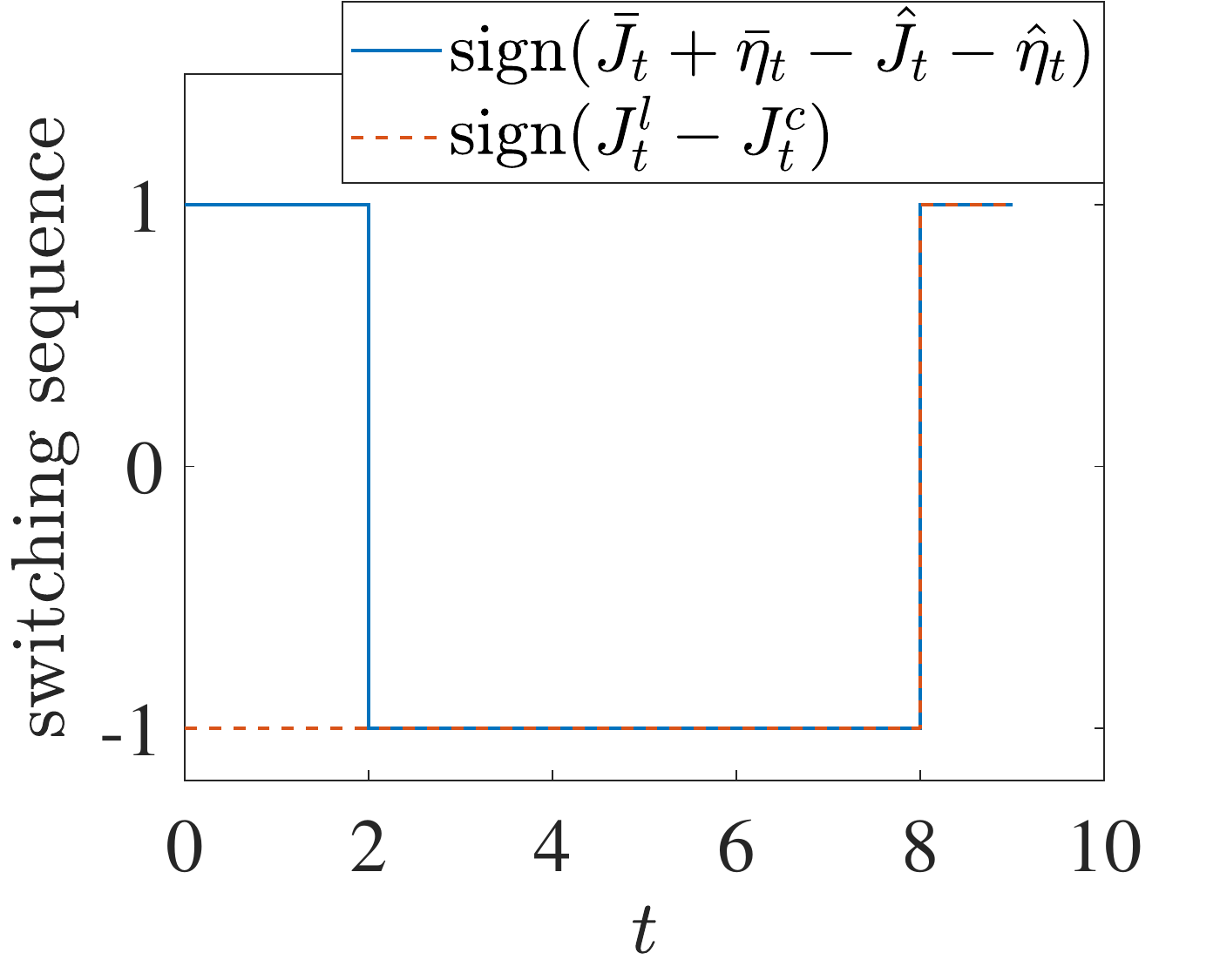}
	}
	\caption{Simulation results of Example 1: (a) Cost-to-go values and their estimates, (b) Switching sequences.}
	\label{fig_J_switch_simple}
\end{figure}

Fig.~\ref{fig_J_switch_simple} illustrates the actual cost-to-go values $J^l_t, J^c_t$, their worst-case estimates $\bar{J}_t + \bar{\eta}_t, \hat{J}_t + \hat{\eta}_t$, and the cloud-local switching sequence. Recall that in the constrained version of our switching policy, \eqref{eqn:policy_2}, which control to apply is determined by two conditions. We have found that the second condition,  $|\hat{x}_t - x_t| \le \delta_t$, is always satisfied in this example, and in this case switching is determined by the sign of $(\bar{J}_t + \bar{\eta}_t)-(\hat{J}_t + \hat{\eta}_t)$. In particular, $\text{sign}(\bar{J}_t + \bar{\eta}_t - \hat{J}_t - \hat{\eta}_t) = 1$ indicates that the cloud MPC control $\hat{u}_t$ is applied and $\text{sign}(\bar{J}_t + \bar{\eta}_t - \hat{J}_t - \hat{\eta}_t) = -1$ indicates that the local MPC control $\bar{u}_{t|t}$ is applied. As discussed in the last paragragh of Section~\ref{sec:policy_1}, our switching policy based on the worst-case cost-to-go estimates $\bar{J}_t + \bar{\eta}_t$ and $\hat{J}_t + \hat{\eta}_t$ may not be the optimal policy. The optimal policy that minimizes the actual cost-to-go can be determined by the sign of $J^l_t - J^c_t$. Nonetheless, the switching sequence determined by our policy matches the optimal switching sequence for $80\%$ of the time instants, indicating that our policy is a reasonably good approximation of the optimal policy. Note that $J^l_t$ and $J^c_t$, which determines the optimal policy, cannot be pre-computed but can only be computed after the actual state trajectories from current time $t$ to terminal time $N$ corresponding to cloud and local MPC implementations have been revealed.

\begin{figure}[!htbp]
	\centering
		\includegraphics[width=2.1 in]{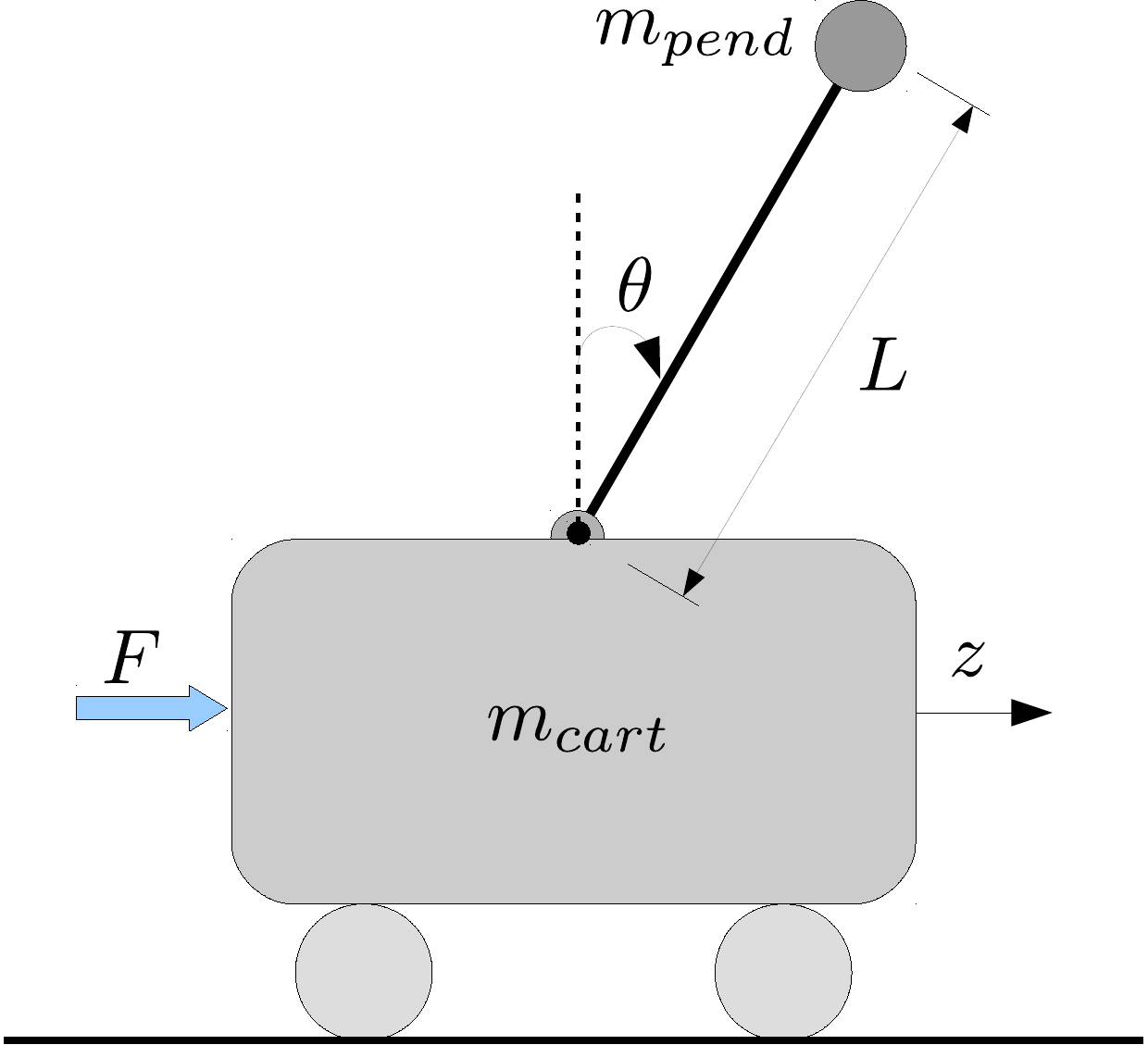}
	\caption{Inverted pendulum on a cart.}
	\label{fig_pendulum_model}
\end{figure}

\subsection{Example 2: Inverted pendulum on a cart} As shown in Fig.~\ref{fig_pendulum_model}, this example considers an inverted pendulum mounted to a cart with its continuous-time model given by \cite{gurumoorthy1993ACC}:
\begin{equation} \label{inverted pendulum}
\begin{aligned}
\ddot{z} &= \frac{F - K_{d}\dot{z} - m_{pend}L\dot{\theta}^{2}\sin(\theta) + m_{pend}g\sin(\theta)\cos(\theta)}{m_{cart} + m_{pend}\sin^{2}(\theta)},
\\
\ddot{\theta} &= \frac{\ddot{z}\cos(\theta)+g\sin(\theta)}{L},
\end{aligned}
\end{equation}
where $z$ is the cart position, $\theta$ is the pendulum angle, $m_{cart} = 1$~kg is the cart mass, $m_{pend} = 1$~kg is the pendulum mass, $L = 0.5$~m is the length of the pendulum, $K_{d} = 10$~Ns/m is the damping parameter, and $g = 9.81$~m/s$^2$ is the gravity acceleration. The system is controlled by a variable force $F$. We discretize the continuous-time model \eqref{inverted pendulum} with a sampling time of $\Delta T = 0.1$~s, and the derived discrete-time model is used by the cloud MPC to compute the control sequence. The lower-fidelity model used by the local MPC is obtained by linearizing \eqref{inverted pendulum} around the equilibrium $\begin{bmatrix}
z & \dot{z} & \theta & \dot{\theta} \end{bmatrix}^{\top}=\begin{bmatrix}
0 & 0 & 0 & 0
\end{bmatrix}^{\top}$. The cost function is selected as
\begin{equation}
J = \sum_{t=0}^{N-1} \left(\|x_{t}\|_{Q} + \|u_{t}\|_{R}\right) + \|x_{N}\|_{Q},
\end{equation}
where $x_{t} = \begin{bmatrix}
z_{t} & \dot{z}_{t} & \theta_{t} & \dot{\theta_{t}}
\end{bmatrix}^{\top}$, $u_{t} = F_{t}$, $Q = \text{diag}(3, 0.4, 3, 0.4)$, and $R = 1 \times 10^{-5}$.

\begin{figure}[!htbp]
	\centering
	\hspace{-0.15 in}
	\subfigure[] {\label{fig_z_pendulum}
		\includegraphics[width=1.76 in]{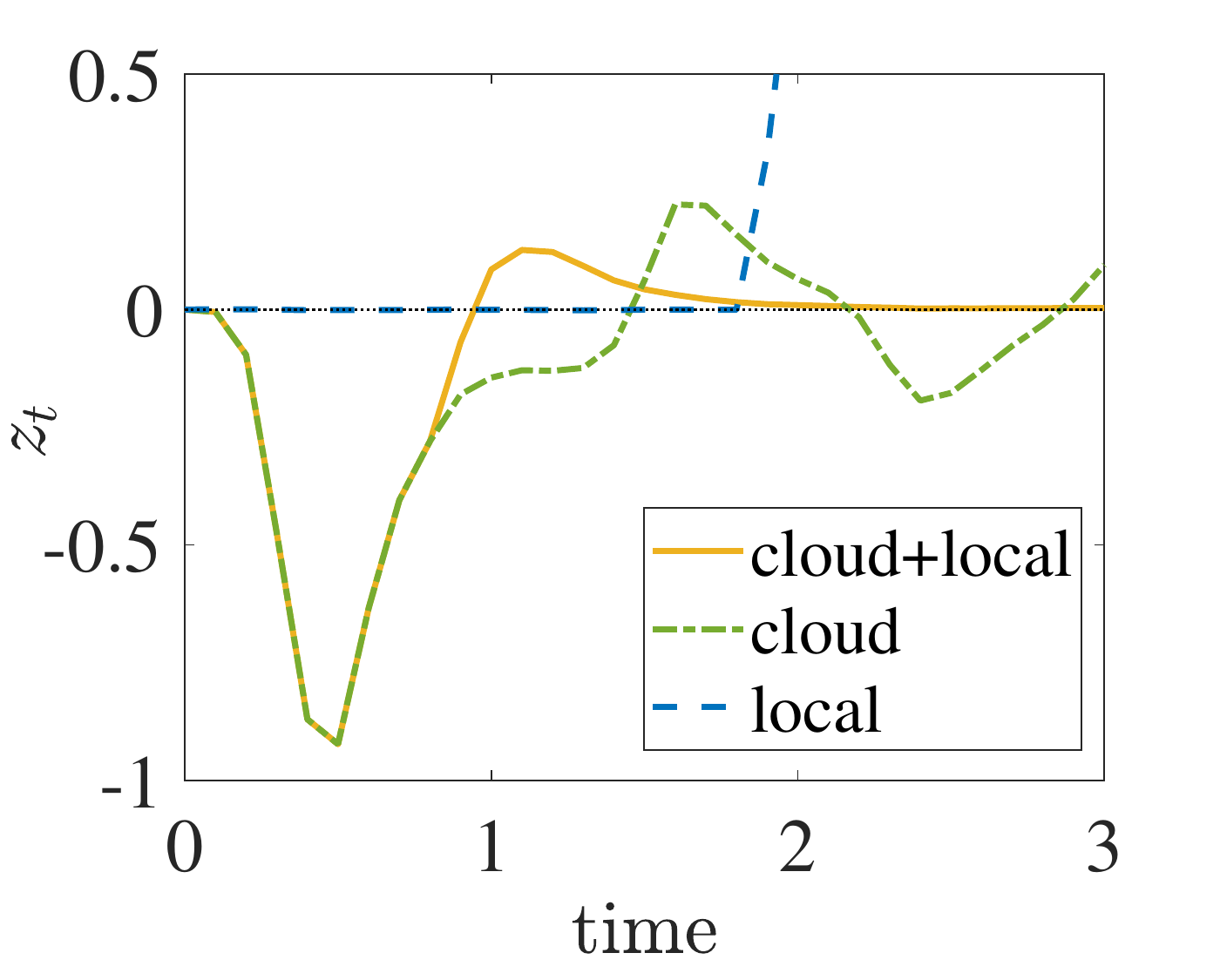}
	}
	\hspace{-0.3 in}
    \subfigure[] {\label{fig_theta_pendulum}
	    \includegraphics[width=1.76 in]{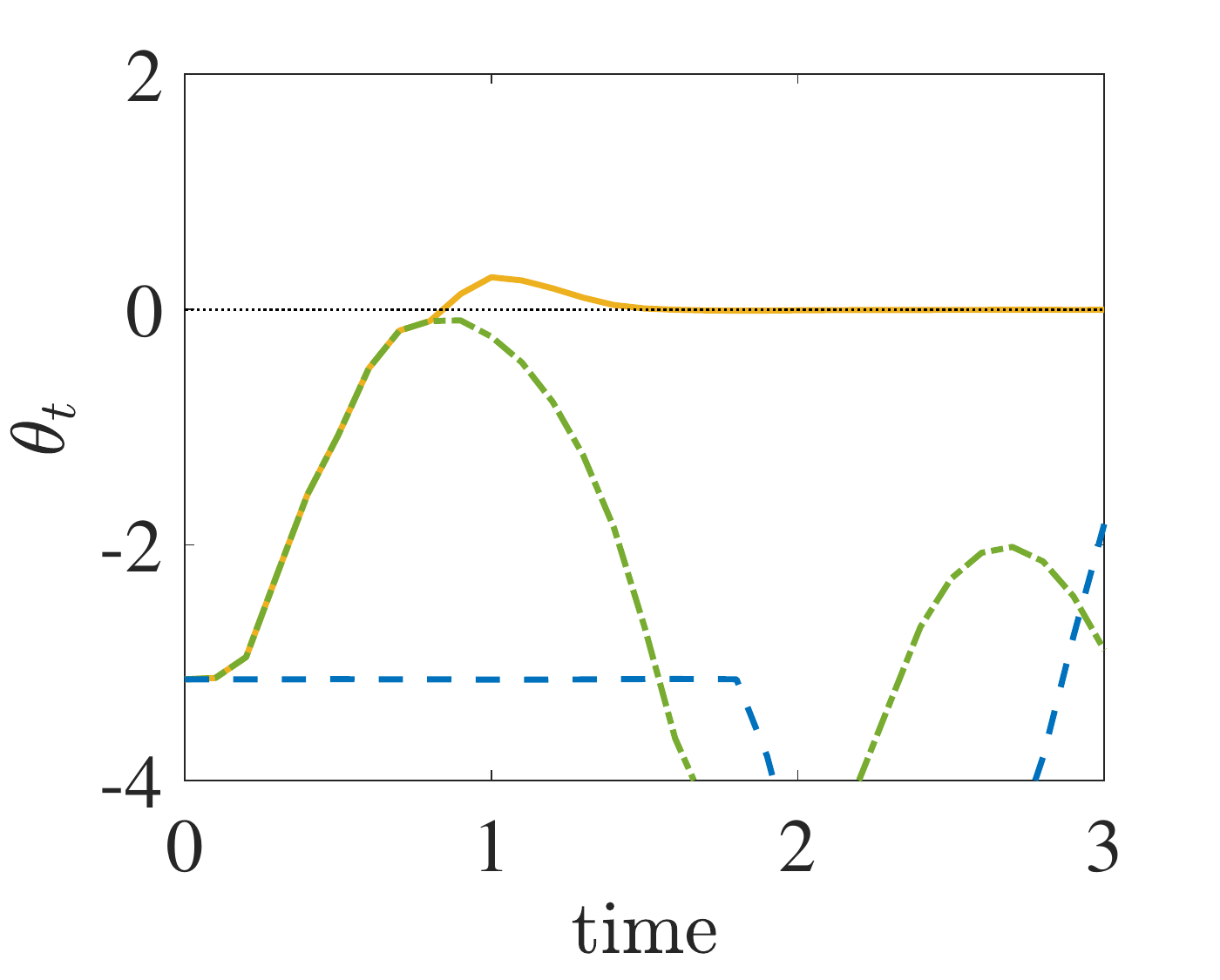}
    }
	\caption{Simulation results of Example 2: State trajectories.}
	\label{fig_pendulum}
\end{figure}
\begin{figure}[!htbp]
	\centering
		\includegraphics[width=2.5 in]{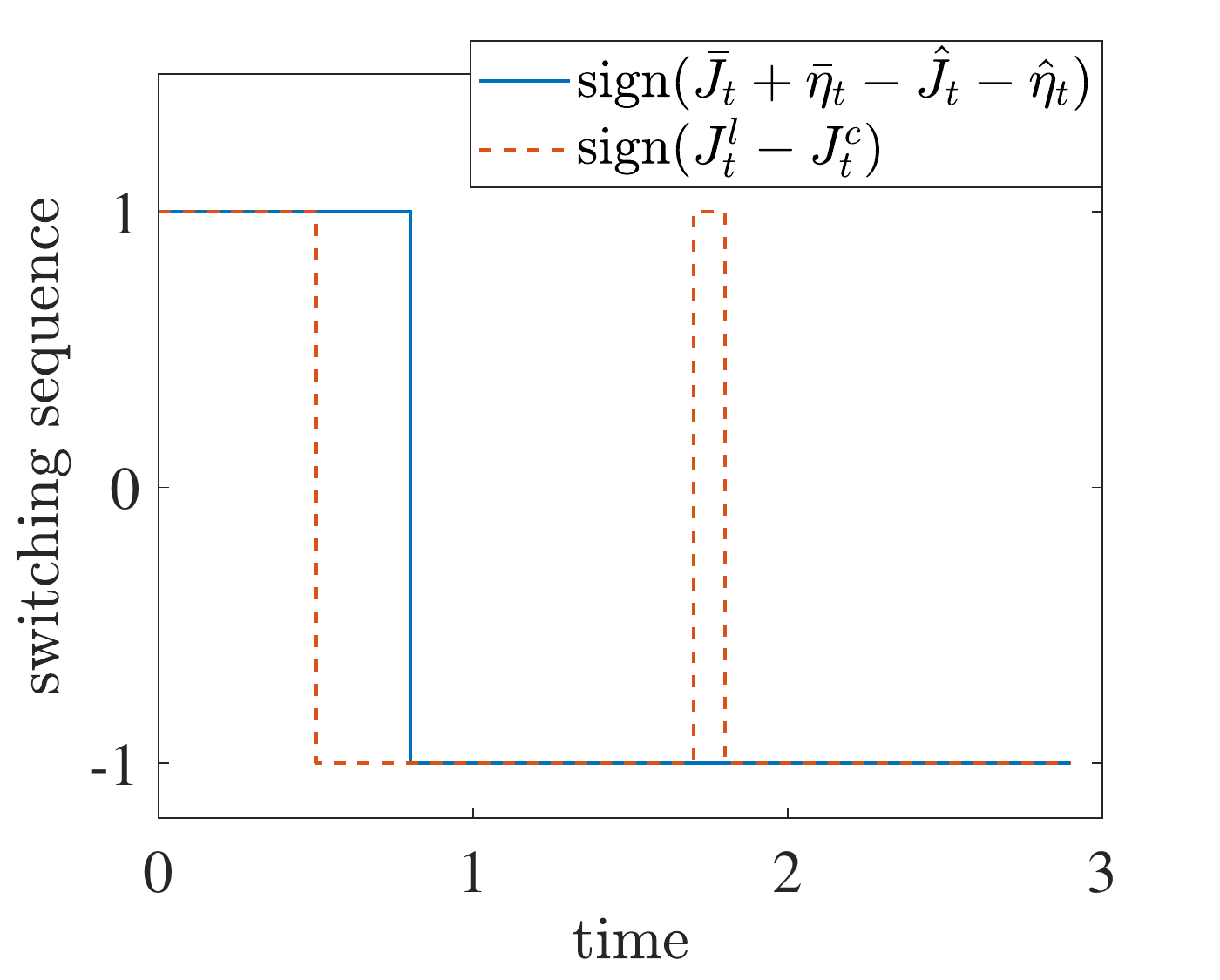}
	\caption{Simulation results of Example 2: Switching sequences.}
	\label{fig_switch_pendulum}
\end{figure}

The system state trajectories corresponding to different MPC schemes are presented in Fig.~\ref{fig_pendulum}, and it can be seen that only our approach of combined cloud + local MPC achieves satisfactory convergence. Recall that the local MPC relies on a linearized model, which is accurate only for states within a small neighborhood of the linearization point. Because the initial condition $x_0$ is far from the linearization point, the local MPC fails to stabilize the inverted pendulum. Similarly, the cloud MPC also fails to stabilize the inverted pendulum due to an accumulation of errors over time. Recall that the cloud MPC, with its control trajectory computed only at the initial time $t = 0$, is essentially an open-loop control scheme, and therefor cannot counteract such errors through feedback. Fig.~\ref{fig_switch_pendulum} shows the switching sequences of $(\bar{J}_{t} + \bar{\eta}_{t}) - (\hat{J}_{t} + \hat{\eta}_{t})$ and $J^{l}_{t}-J^{c}_{t}$ with a 86.67\% match.

\begin{figure}[!htbp]
	\centering
		\includegraphics[width=2.4 in]{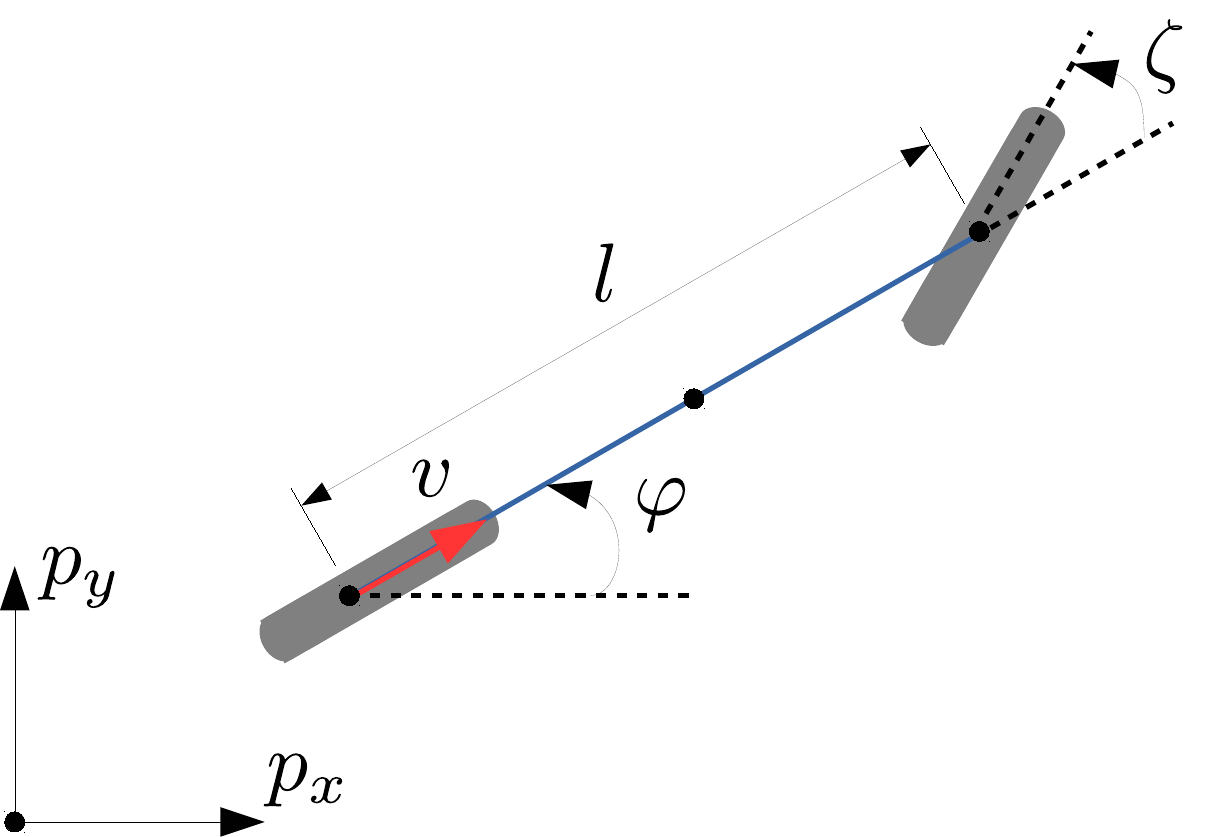}
	\caption{Kinematic bicycle model.}
	\label{fig_bicycle_model}
\end{figure}

\subsection{Example 3: Autonomous vehicle path following} This example considers the path following of an autonomous vehicle as presented in Fig.~\ref{fig_bicycle_model}. The kinematic model of the vehicle is described by \cite{DeLuca1998}:
\begin{equation} \label{vehicle model}
\begin{aligned}
\dot{p}_{x} &= v\cos(\varphi),
\\
\dot{p}_{y} &= v\sin(\varphi),
\\
\dot{\varphi} &= \frac{v}{l}\tan(\zeta),
\end{aligned}
\end{equation}
where $\left(p_x, p_y\right)$ denotes the position of the vehicle, $\varphi$ is the yaw angle, and $l$ is the wheelbase. The speed $v$ and steering angle $\delta$ are the control variables for the vehicle. Denote $x^{\text{ref}} = \begin{bmatrix}
p_{x}^{\text{ref}} & p_{y}^{\text{ref}} & \varphi^{\text{ref}}
\end{bmatrix}^{\top}$ and $u^{\text{ref}} = \begin{bmatrix}
v^{\text{ref}} & \zeta^{\text{ref}}
\end{bmatrix}^{\top}$ as the reference path to follow and its corresponding control variables. Then, by letting $\gamma = \tan(\zeta)$ ($\gamma^{\text{ref}} = \tan(\zeta^{\text{ref}})$), $\tilde{x} = \begin{bmatrix} \tilde{p}_{x} & \tilde{p}_{y} & \tilde{\varphi} \end{bmatrix}^{\top} = \begin{bmatrix} p_{x}-p_{x}^{\text{ref}} & p_{y}-p_{y}^{\text{ref}} & \varphi-\varphi^{\text{ref}} \end{bmatrix}^{\top}$, and $\tilde{u} = \begin{bmatrix} \tilde{v} & \tilde{\gamma} \end{bmatrix}^{\top} = \begin{bmatrix} v-v^{\text{ref}} & \gamma-\gamma^{\text{ref}} \end{bmatrix}^{\top}$, the model \eqref{vehicle model} can be written as
\begin{equation} \label{error model}
    \dot{\tilde{x}} = A \tilde{x} + B \tilde{u} + f,
\end{equation}
where
\[
\begin{aligned}
& A = \begin{bmatrix}0 & 0 & -v^{\text{ref}}\sin(\varphi^{\text{ref}}) \\ 0 & 0 &  v^{\text{ref}}\cos(\varphi^{\text{ref}}) \\ 0 & 0 & 0 \end{bmatrix}, B = \begin{bmatrix} \cos(\varphi^{\text{ref}}) & 0 \\ \sin(\varphi^{\text{ref}}) & 0 \\ \frac{\gamma^{\text{ref}}}{l} & \frac{v^{\text{ref}}}{l} \end{bmatrix},
\\[4pt]
& f \!=\!\small{\begin{bmatrix}
\left(\cos(\tilde{\varphi} \!+\! \varphi^{\text{ref}})\!-\!\cos(\varphi^{\text{ref}})\right)(\tilde{v} \!+\! v^{\text{ref}}) \!+\! v^{\text{ref}}\sin(\varphi^{\text{ref}})\tilde{\varphi}
\\
\left(\sin(\tilde{\varphi}\!+\!\varphi^{\text{ref}})\!-\!\sin(\varphi^{\text{ref}})\right)(\tilde{v}\!+\!v^{\text{ref}}) \!-\! v^{\text{ref}}\cos(\varphi^{\text{ref}})\tilde{\varphi}
\\
\frac{\tilde{v}\tilde{\gamma}}{l}
\end{bmatrix}}.
\end{aligned}
\]
The continuous-time model \eqref{error model} is discretized with a sampling time of $\Delta T = 0.05$~s as the higher-fidelity model for cloud MPC, while the linear model used by the local MPC is obtained by neglecting the nonlinear term $f$. Furthermore, for tracking the reference path, the cost function is design as
\begin{equation}
\begin{aligned}
J &= \sum_{t=0}^{N-1} \left( \| \tilde{x}_{t} \|_{Q} + \|\tilde{u}_{t}\|_{R} \right)
+ \|\tilde{x}_{N}\|_{Q},
\end{aligned}
\end{equation} 
where $Q = \text{diag}(3, 3, 0.01)$ and $R = 0.001 I_{2 \times 2}$. Note that the system parameters $A$, $B$, and $f$ are related to the time-varying signal $u^{\text{ref}}$. In this case, we derive the constants $a$, $L_{f}$, and $M_{f}$ for cost estimation according to Remark~5.

\begin{figure}[!htbp]
	\centering
	\includegraphics[width=3.1 in]{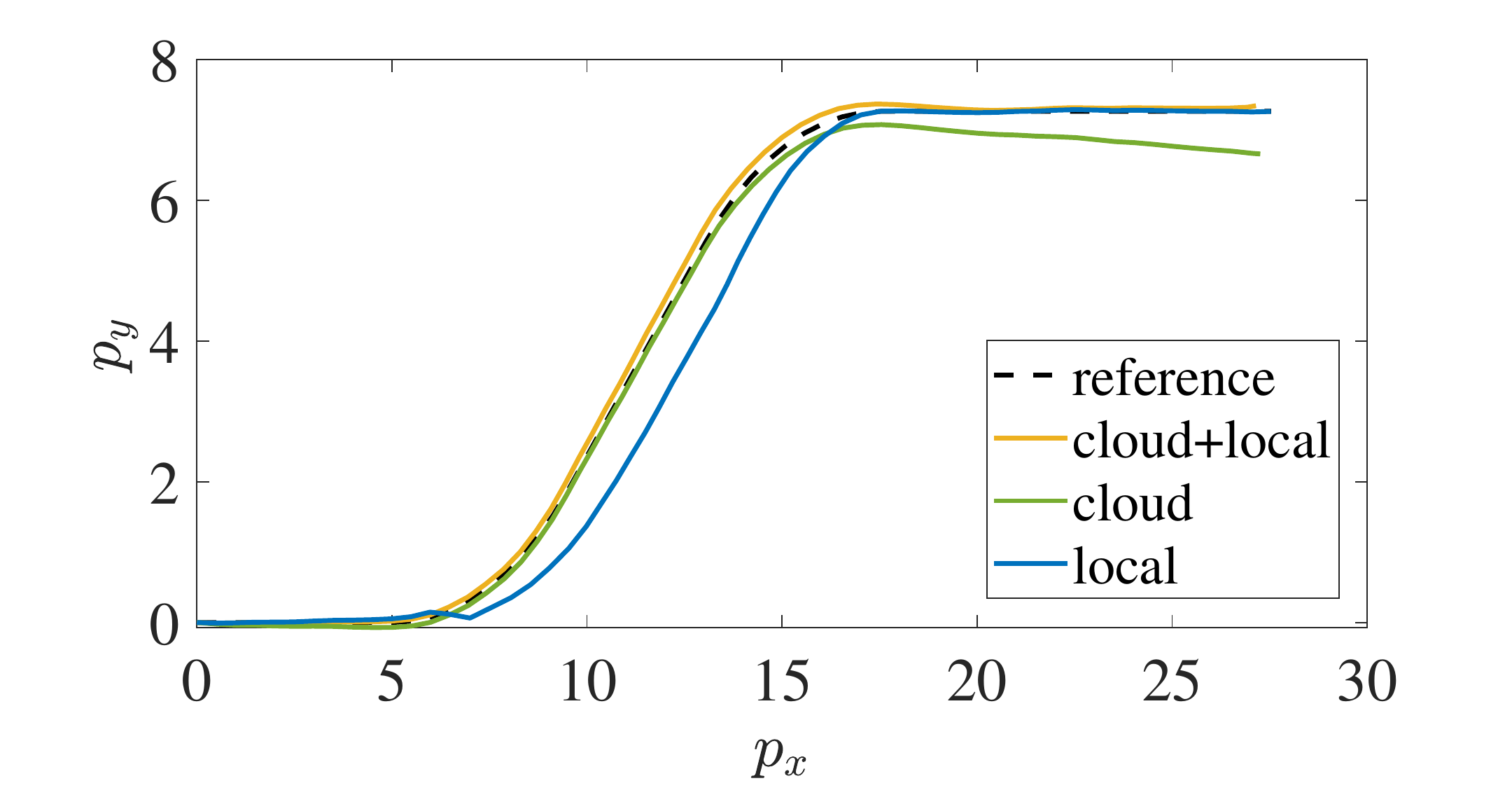}
	\caption{Simulation results of Example 3: Position trajectories.}
	\label{fig_vehicle}
\end{figure}

\begin{figure}[!htbp]
	\centering
	\includegraphics[width=2.5 in]{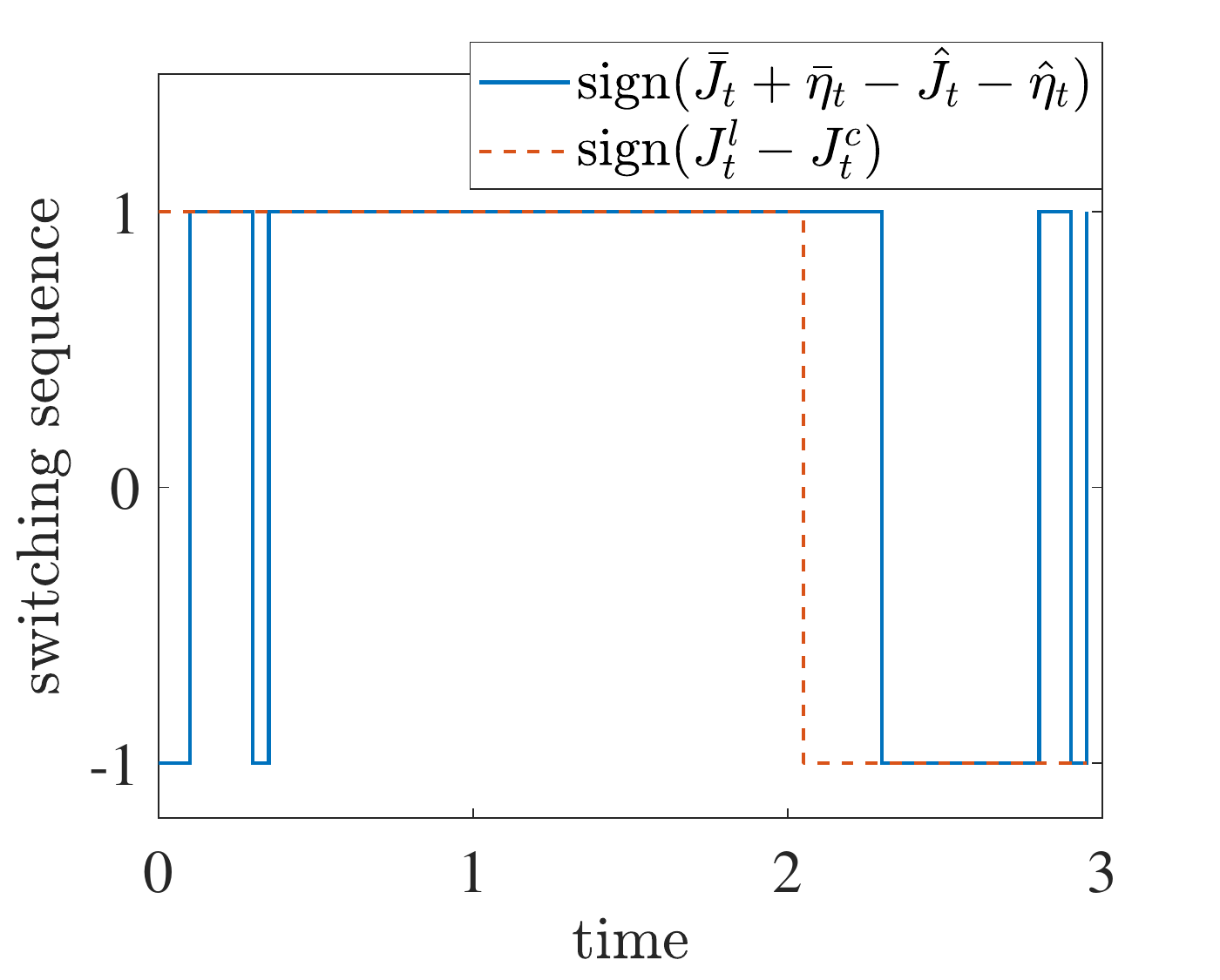}
	\caption{Simulation results of Example 3: Switching sequences.}
	\label{fig_switch_vehicle}
\end{figure}

The position trajectories of the vehicle corresponding to different MPC methods are illustrated in Fig.~\ref{fig_vehicle}, and it is clear that our cloud  + local MPC approach achieves the most accurate path following. Furthermore, Fig.~\ref{fig_switch_vehicle} presents the switching sequences of $(\bar{J}_{t} + \bar{\eta}_{t}) - (\hat{J}_{t}+\hat{\eta}_{t})$ and $J^{l}_{t} - J^{c}_{t}$ with a 81.67\% match.

\section{Conclusions and Future Work}\label{sec:conclusion}

In this paper, we developed a cloud-assisted model predictive control (MPC) framework for finite-duration control tasks. In this framework, cloud-computed control trajectory based on a higher-fidelity nonlinear model and local shrinking-horizon MPC solutions based on a lower-fidelity linear model are fused by a switching policy to achieve improved control performance in the presence of cloud/local model prediction errors (due to plant-model mismatches and cloud request-response/communication delay effects). We analyzed properties of the proposed cloud-assisted MPC framework and established approaches to robustly handling constraints within this framework in spite of model prediction errors. We then demonstrated the effectiveness of the framework in terms of improving overall control performance using multiple simulation examples, including an automotive control example to illustrate its potential industrial applications.

For future work, we will investigate methods to refine our switching policy as well as investigate other fusion schemes to achieve further improved control performance. We will also extend our cloud-assisted MPC framework from finite-duration control tasks to a broader range of application scenarios.

% \addtolength{\textheight}{-12cm}  % This command serves to balance the column lengths
                                  % on the last page of the document manually. It shortens
                                  % the textheight of the last page by a suitable amount.
                                  % This command does not take effect until the next page
                                  % so it should come on the page before the last. Make
                                  % sure that you do not shorten the textheight too much.

%\bibliographystyle{ieeetr}
\bibliographystyle{IEEEtran}
\bibliography{IEEEfull,ref}

\begin{biography}[{\includegraphics[width=1in,height=1.25in,clip,keepaspectratio]{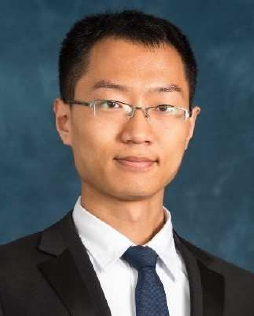}}]{Nan Li}
		received the Ph.D. degree in aerospace engineering and the M.S. degree in mathematics from the University of Michigan, Ann Arbor, MI, USA, in early 2021 and 2020, respectively. He is currently a postdoctoral research fellow there. 
		
		His research interests are in stochastic control, game theory, multi-agent systems, and safety-critical systems.
\end{biography}

\begin{biography}[{\includegraphics[width=1in,clip,keepaspectratio]{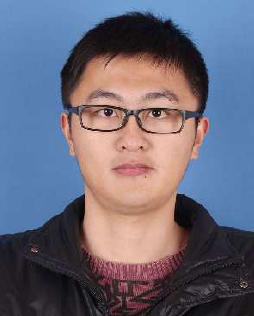}}]{Kaixiang Zhang}
		received the B.E. degree in automation from Xiamen University, Xiamen, China, in 2014, and the Ph.D. degree in control science and engineering from Zhejiang University, Hangzhou, China, in 2019.

		He currently holds a postdoctoral position with the Department of Mechanical Engineering at Michigan State University. His research interests include visual servoing, robotics, and nonlinear control.
\end{biography}

\begin{biography}[{\includegraphics[width=1in,height=1.25in,clip,keepaspectratio]{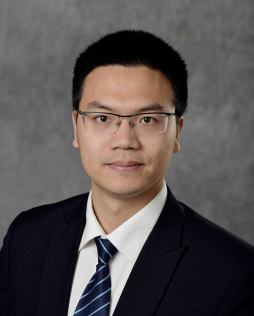}}]{Zhaojian Li}
    	received his B. Eng. degree from Nanjing University of Aeronautics and Astronautics in 2010. He obtained M.S. (2013) and Ph.D. (2015) in Aerospace Engineering (flight dynamics and control) at the University of Michigan, Ann Arbor. 
    	
    	He is currently an Assistant Professor with the department of Mechanical Engineering at Michigan State University. His research interests include learning-based control, nonlinear and complex systems, and robotics and automated vehicles. He is a recipient of NSF CAREER Award.
\end{biography}

\begin{biography}[{\includegraphics[width=1in,height=1.25in,clip,keepaspectratio]{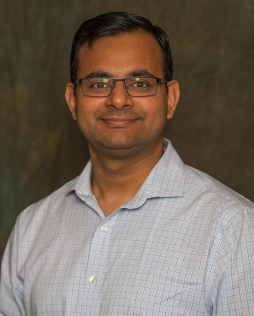}}]{Vaibhav Srivastava}
        received the B.Tech. degree (2007) in mechanical engineering from the Indian Institute of Technology Bombay, Mumbai, India; the M.S. degree in mechanical engineering (2011), the M.A. degree in statistics (2012), and the Ph.D. degree in mechanical engineering (2012) from the University of California at Santa Barbara, Santa Barbara, CA. He served as a Lecturer and Associate Research Scholar with the Mechanical and Aerospace Engineering Department, Princeton University, Princeton, NJ from 2013-2016. 
        He is currently an Assistant Professor with the Electrical and Computer Engineering at Michigan State University. He is also affiliated with Mechanical Engineering, Cognitive Science Program, and Connected and Autonomous Networked Vehicles for Active Safety (CANVAS) Program.  His research focuses on Cyber Physical Human Systems with emphasis on mixed human-robot systems, networked multi-agent systems, aerial robotics, and connected and autonomous vehicles.
\end{biography}

\begin{biography}[{\includegraphics[width=1in,height=1.25in,clip,keepaspectratio]{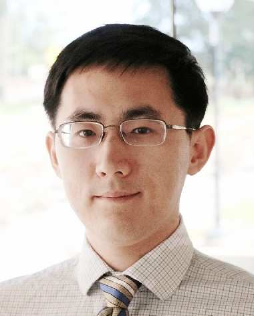}}]{Xiang Yin}
        was born in Anhui, China, in 1991. He received the B.Eng degree from Zhejiang University in 2012, the M.S. degree from the University of Michigan, Ann Arbor, in 2013, and the Ph.D degree from the University of Michigan, Ann Arbor, in 2017, all in electrical engineering. Since 2017, he has been with the Department of Automation, Shanghai Jiao Tong University, where he is an Associate Professor. His research interests include formal methods,  discrete-event systems and cyber-physical systems.

        Dr.\ Yin also received the IEEE Conference on Decision and Control (CDC) Best Student Paper Award Finalist in 2016. He is the co-chair of the IEEE CSS Technical Committee on Discrete Event Systems. He is also a member of the IEEE CSS Conference Editorial Board.
\end{biography}

\end{document}